\documentclass[11pt]{article}
\usepackage{amsthm,amssymb,eucal,nath}
\usepackage{algorithm,algorithmic}
\newtheorem{theorem}{Theorem}[section]
\newtheorem{proposition}[theorem]{Proposition}
\newtheorem{lemma}[theorem]{Lemma}
\newtheorem{corollary}[theorem]{Corollary}
\newtheorem{construction}[theorem]{Construction}
\newtheorem{remark}[theorem]{Remark}
\newtheorem{example}[theorem]{Example}

\newtheorem{definition}[theorem]{Definition}

\algsetup{linenodelimiter=\ }

\def\eqref#1{{\rm(\ref{#1})}}

\def\picbdot{\makebox(0,0){\hbox{$\bullet$}}}

\def\LCM{\mathop{\mathrm{lcm}}\nolimits}

\begin{document}

\title{Sufficient set of integrability conditions of an orthonomic system}
\author{\small M. Marvan,
\\\footnotesize Mathematical Institute of the Silesian University at Opava,
\\\footnotesize Na Rybn\'\i\v{c}ku 1, 746 01 Opava, Czech Republic}
\date{}
\maketitle

\begin{abstract}
Every orthonomic system of partial differential equations is known to possess a finite number of integrability conditions sufficient to ensure the validity of all. 
Herewith we offer an efficient algorithm to construct a sufficient set of integrability conditions free of redundancies.
\end{abstract}

\section{Introduction}

\paragraph{1.} 
Somewhat surprisingly, existing literature on orthonomic systems lacks an effective construction of a provably irredundant sufficient set of integrability conditions. The aim of this paper is to fill this gap.

A subproblem attracted much attention in polynomial elimination theory. When computing Gr\"obner bases through the famous Buchberger algorithm~\cite{Buch}, $S$-polynomials arise as analogues of integrability conditions.
The problem of minimizing the number of $S$-polynomials was already addressed by Buchberger~\cite{Buch-1}, with many later developments (\cite{W-B-L-R,G-M,C-K-R,A-H}).
The basic idea, exploiting syzygies, later migrated back to differential algebra (Boulier~\cite{Bou}) and Riquier theory (Reid's school~\cite{Ru,R-R-W}).
Within the syzygy approach one detects superfluous ``critical pairs'' ($S$-polynomials or integrability conditions) and removes them sequentially. 
In practical implementations such as Witt\-kopf's dissertation~\cite{Witt} detection of superfluous integrability conditions turns out to be nearly exhaustive (see examples at the end of the paper). 
The solution presented in this paper is of rather different nature and amounts to a direct construction of the irredundnat set of nontrivial integrability conditions, making explicit all remaining freedom of choice.

\paragraph{2.} 
Present developments of formal integrability theory are, to a great extent, driven by computer algebra applications, especially solution of large systems of overdetermined PDE connected with computation of symmetries, conservation laws, and other invariants of PDE (see surveys \cite{He,CRC}). Since input systems consisting of hundreds of equations are not uncommon, efficiency of the algorithms is an important issue.

Initially (Riquier~\cite{Riq}, Janet~\cite{Jan}) the basic question was which coefficients of Taylor expansion of a solution could be chosen arbitrarily (parametric derivatives) and which were then uniquely determined by the system (principal derivatives). 
As is well known, hidden dependences between parametric derivatives lead to integrability (or compatibility) conditions. A system with or without unsatisfied integrability conditions is said to be {\it active} or {\it passive,} respectively.
The procedure of augmenting an active system with its integrability conditions is called the {\it completion.} The augmented system is not necessarily passive, since new integrability conditions can emerge. However, repeated completion is guaranteed to stop after a finite number of steps under fairly general assumptions (Tresse~\cite{Tr}; see \cite{R-L-W} for an overview). 

Conventional wisdom says that computing integrability conditions amounts to taking cross-derivatives. But the notion of cross-derivative applies only to orthonomic systems (ones resolved with respect to ``highest'' derivatives). Moreover, integrability conditions can depend substantially on the way the system is resolved (as opposed to the Cartan and Spencer geometric theory of involutivity~\cite{Pom} and the recent theory of Mayer brackets~\cite{Kr-LyI,Kr-LyII}).  
On the other hand, if we accept all the unpleasant consequences, as we do in the present paper, we find ourselves placed in an environment tailored for easy and efficient implementation of reduction. Reduction is a procedure to compute a normal form modulo identifications following from the system.

\paragraph{3.} 
As usual, this paper deals with the infinitely prolonged system $\Sigma^\infty$, which consists of individual equations of the input system $\Sigma$ differentiated with respect to every combination of independent variables. Under a suitable ranking of derivatives, every equation of $\Sigma^\infty$ comes out resolved with respect to a principal derivative, which substantially simplifies the procedure of reduction with respect to $\Sigma^\infty$. The main technical difficulty is that reduction is not unique unless the system $\Sigma$ is passive, which is not guaranteed before the completion algorithm is finished.
To circumvent this problem, we consider a triangular subsystem $\Sigma'$ of $\Sigma^\infty$ with the goal to show equivalence of $\Sigma'$ and $\Sigma^\infty$. 

Our main result says how to locate nonignorable integrability conditions in the monomial ideal(s) generated by the system. Namely, we determine substantial integrability conditions at a principal derivative $u^k_\mu$ from connected components of certain subset $\mathcal X^k_\mu$ of the monomial ideal ordered by divisibility. The idea bears certain surprising resemblance to that of ``subconnectedness'' (see~\cite{W-Bu,Kuch,Win} and references therein) having roots in Gr\"obner basis theory as well, although a closer look reveals substantial differences. 

It is easily seen that $u^k_\mu$ must be a cross-derivative in order to possess nontrivial integrability conditions.
This observation immediately leads to a canonical and transparent construction of a sufficient set of integrability conditions (Construction~\ref{suffcc}). Next, this set is shown to be irredundant in a sense that no integrability condition can be ignored. 
Considering the ``staircase diagram'' associated with the monomial ideal, the main result of this paper provides identification of the vertices where nontrivial integrability conditions reside. This opens the door to asking and answering various combinatorial questions, which is however out of the scope of the present paper.

It is often argued that orthonomic essentially means linear from the practical perspective, since arbitrary nonlinearities can occur at later stages of completion. Let us stress that results of this paper rather loosely depend on orthonomicity, since for the integrability conditions to show up it is not necessary that the system be explicitly resolved. Computation of derivatives of implicit functions being little problem, difficulties lie with implementing reduction. In case of polynomially nonlinear systems a help comes from the theory of triangular systems (see~\cite{Hu} for an overview). That said, we leave this issue to a further study. 

An extended abstract of the previous version of this paper appeared in Proceedings of the GIFT 2006 conference~\cite{GIFT}. Another abridged exposition will be made through the book~\cite{V-K-K-M}.

\section{Orthonomic systems}
\label{sect:orth}

In this section we recall standard facts and fix our notation.
We denote by $\mathcal U = \{u^1,\dots,u^l\}$ a set of dependent variables and by $\mathcal X = \{x_1,\dots,x_n\}$ a set of independent variables. 
Consider the free commutative monoid $\mathcal X^*$ over $\mathcal X$.
An arbitrary element $\mu \in \mathcal X^*$ is of the form $\mu = x_1^{r_1} \dots x_n^{r_n}$, $r_1 \ge 0, \dots,\ r_n \ge 0$, and will be called a {\it Janet monomial\/}~{\cite{Jan}}. 
A derivative
$\partial^{r_1 + \dots + r_n} u^k/\partial x_1^{r_1} \dots \partial x_n^{r_n}$
can be identified with a pair $(u^k, x_1^{r_1} \dots x_n^{r_n}) \in \mathcal U \times \mathcal X^*$.
It will be convenient to denote derivatives as $u^k_\mu$, $\mu \in \mathcal X^*$. 
Dependent variables $u^k$ can be idetified with derivatives $u^k_1$ of order~$0$.

Elements of $\mathcal X \cup (\mathcal U \times \mathcal X^*)$ bijectively correspond to local coordinates on an appropriate infinite-dimensional jet space $J^\infty$~\cite{B-C-D-K-K-S-T-V-V,Pom,Sau}. 
For the purpose of understanding the present paper it is sufficient to think of $J^\infty$ as an infinite-dimensional space equipped with coordinates indexed by elements of $\mathcal X \cup (\mathcal U \times \mathcal X^*)$.
Smooth functions are defined as mappings $J^\infty \to \mathbb R$ that (locally) depend on only a finite number of coordinates.
For each $x \in \mathcal X$, the {\it total derivative}
$$
D_x = \frac\partial{\partial x} + \sum_{k,\mu} u^k_{\mu x} \frac\partial{\partial u^k_\mu}.
$$
can be viewed as a vector field on $J^\infty$ (or differentiation of the $\mathbb R$-algebra of smooth functions on $J^\infty$).
Observe that $D_x$ acts on a derivative $u^k_\mu$ by multiplying the Janet monomial $\mu$ by~$x$. As is well known, total derivatives commute. For every $\nu \in \mathcal X^*$ the corresponding composition of total derivatives is denoted by $D_\nu$.

\begin{definition} \label{higher}
\rm
We denote by $\le$ the relation of divisibility of monomials and by
$\mu/\nu$ the quotient of monomials $\mu,\nu \in \mathcal X^*$ whenever $\mu \ge \nu$.
If so, then the derivative $u^k_\nu$ is said to be {\it lower} than the derivative $u^k_\mu$. We say that $\nu$ is {\it strictly lower} than $\mu$, writing $\nu < \mu$, if $\nu \le \mu$ and $\nu \ne \mu$.  
\end{definition}

Essential in Riquier's theory is a suitable ordering of derivatives compatible with differentiation.

\begin{definition} \rm
\label{def:adm}
A {\it ranking\/} is a linear ordering $\preceq$ of the set $\mathcal U \times \mathcal X^*$ of derivatives such that
$$
p \prec D_x p,
\\
p \prec p' \Rightarrow D_x p \prec D_x p'
$$
for all $p,p' \in \mathcal U \times \mathcal X^*$ and every $x \in \mathcal X$.
\end{definition}

Obviously, we have the implication $q \le p \Rightarrow q \preceq p$, but not the converse.
It is an easy consequence of Dickson's lemma (see also Thomas~\cite{Th}) that $\mathcal U \times \mathcal X^*$ is a well-ordered set, in particular, every decreasing chain is finite.
This enables transfinite induction through the set $\mathcal U \times \mathcal X^*$. 
A complete classification of rankings has been obtained by Rust~\cite{Ru,Ru-R}.

Now we take into consideration a system $\Sigma$ of finitely many partial differential equations, resolved with respect to diverse derivatives
\begin{equation}
\label{eq:Sigma}
u^k_\mu = \Phi^k_\mu,
\end{equation}
where $\Phi^k_\mu$ are smooth functions on $J^\infty$.
By $`dom \Sigma$ we denote the set of all derivatives, appearing on the left-hand side of at least one of the equations \eqref{eq:Sigma}
(even though $`dom \Sigma$ is determined by the form~\eqref{eq:Sigma} rather than by $\Sigma$ itself).

The basic object of interest in formal integrability theory is the associated {\it infinitely prolonged\/} system $\Sigma^\infty$ consisting of all possible differential consequences:
$$
\numbered\label{eq:prol}
u^k_{\mu\nu} = D_\nu \Phi^k_\mu, \qquad \nu \in \mathcal X^*.
$$
Derivatives $u^k_\sigma$ appearing on the left-hand side of some equation from $\Sigma^\infty$, i.e., belonging to $`dom \Sigma^\infty$, are said to be {\it principal}. The other derivatives are said to be {\it parametric}.
Thus, a derivative is principal if it either belongs to $`dom\Sigma$ or is a derivative of some derivative from $`dom\Sigma$.

Now, each equation of the input system $\Sigma$ is supposed to be resolved with respect to the highest rank derivative it contains, and the right-hand sides are supposed to be free of principal derivatives.  
Summing up, the system~\eqref{eq:Sigma} is assumed to be orthonomic in the following sense:

\begin{definition} \rm
\label{1-1-n}
A system of equations $\Sigma$ in the form \eqref{eq:Sigma} is said to be 

\paritem{--} {\it triangular}, if for every derivative $q \in `dom\Sigma$ there is exactly one equation with $q$ appearing on its left-hand side;

\paritem{--} {\it normal} with respect to a ranking $\preceq$, or $\preceq$-normal, if every derivative $q \in `dom\Sigma$ is $\preceq$-maximal in its equation;

\paritem{--} {\it autoreduced}, if no principal derivative occurs on the right-hand side of any equation;

\paritem{--} {\it orthonomic} with respect to a ranking $\preceq$, or $\preceq$-orthonomic, if it is triangular, $\preceq$-normal, and autoreduced.
\end{definition}

The following easy observation is crucial to Riquier's theory: The infinitely prolonged system $\Sigma^\infty$ is normal when $\Sigma$ is normal.
However, the property of being triangular is usually lost in $\Sigma^\infty$ (which is why integrability conditions occur). Having autoreduced right-hand sides of $\Sigma$ will be useful in the sequel, while for $\Sigma^\infty$ no such property is needed.

We came to the point where division between multiplicative and nonmultiplicative variables~\cite{Jan} enters the discourse in orthodox exposition of the Riquier--Janet theory. Formalized by Gerdt and Blinkov~\cite{Ger,G-B1} (these works triggered an active thread of research in polynomial elimination theory), so-called involutive divisions became a standard tool to prescribe a unique right-hand side to every principal derivative from $`dom \Sigma^\infty$. Selecting a unique right-hand side is convenient, but not absolutely necessary (Reid's school~\cite{Re,R-W-B} succeeds without it).

\section{Reduction subsystem}
\label{sect:rs}

Here we take a another path to uniqueness. Namely, we consider an arbitrary triangular subsystem $\Sigma'$ of $\Sigma^\infty$ with $`dom\Sigma' = `dom\Sigma^\infty$ and call it a {\it reduction subsystem} of $\Sigma^\infty$, since it provides us with a unique reduction. Then completion (see Introduction) may be regarded as a procedure to establish equivalence of $\Sigma'$ and~$\Sigma^\infty$. All algorithms to follow actually refer only to a finite part of $\Sigma'$, hence its infiniteness does not hamper computability.

\begin{construction} \rm \label{1-1sub} 
For each $u^k_\mu \in `dom\Sigma^\infty$, choose arbitrary $\xi \in \mathcal X^*$ such that $u^k_{\mu/\xi} \in `dom\Sigma$ (so that $\Phi^k_{\mu/\xi}$ exists) and put $\Psi^k_\mu = D_\xi\Phi^k_{\mu/\xi}$. 
The triangular system of equations
\begin{equation}
\label{eq:Sigma'}
u^k_\mu = \Psi^k_\mu, \quad u^k_\mu \in `dom\Sigma^\infty
\end{equation}
is the reduction subsystem sought.
\end{construction}

This construction gives us a considerable freedom of choice, measured by the number of elements in $`dom\Sigma$ that are lower than $u^k_\mu$ (see Definition~\ref{higher}).

Given an expression $F$ depending on a finite number of derivatives, one can apply equations of the reduction subsystem \eqref{eq:Sigma'} as substitutions to
obtain an ``equivalent'' expression $SF$ without dependence on principal derivatives. 
In each step, the highest rank principal derivative $p = u^k_\mu$ the expression $F$ actually depends on is substituted by the corresponding expression $\Psi^k_\mu$ from Construction~\ref{1-1sub}. Such steps can be repeated while $F$ depends on principal derivatives. There can be only a finite number of these steps since $\prec$ has the descending chain property. Hence the reduction procedure is algorithmic. Effective implementations are available, see, e.g., Wittkopf~\cite{Witt}. 

Reduction $S$ is an $\mathbb R$-algebra homomorphism $C^\infty J^\infty \to C^\infty J^\infty$ and satisfies $S \circ S = S$.

Applying reduction $S$ on the right-hand sides $\Psi^k_\mu$ of system \eqref{1-1sub} we obtain an autoreduced reduction system (see Definition~\ref{1-1-n}), whose the right-hand sides $\Psi^k_\mu$ depend only on parametric derivatives, hence we have simply $Su^k_\mu = \Psi^k_\mu$.
The autoreduced system generates the same reduction as the unreduced one.
Actually, the previous version of this paper (see the extended abstract~\cite{GIFT}) depended on use of an autoreduced reduction subsystem.  
However, autoreduction is no longer necessary in practical implementations (see Remark~\ref{Buch-1}).

Our main result below (Theorem~\ref{th:pass}) shows that reduction and total derivatives $D_x$ on $J^\infty$ satisfy 
$$
S \circ D_x \circ S = S \circ D_x.
$$
The next example demonstrates that we have no such property until we know that $\Sigma$ is passive.

\begin{example} \rm
A simple example of an active system $\Sigma$ with $S D_x S F \ne S D_x F$ is
$$
u_x = f(u), \quad u_y = g(u).
$$
Let the reduction subsystem $\Sigma'$ contain the equation
$$
u_{xy} = D_y f
$$
rather than its alternative $u_{xy} = D_y g$ (the ranking $\prec$ can be arbitrary).
For $F = u_y$ we obtain
$$
S D_x S u_y = S D_x g = S(\frac{\partial g}{\partial u} u_x) = \frac{\partial g}{\partial u} f, \\
S D_x u_y = S u_{xy} = S D_y f = S(\frac{\partial f}{\partial u} u_y)
 = \frac{\partial f}{\partial u} g.
$$
Observe that $S D_x S u_y = S D_x u_y$ is exactly the integrability condition 
$$
\frac{\partial f}{\partial u} g = \frac{\partial g}{\partial u} f
$$
for the system $\Sigma$. 
\end{example}

\section{Integrability conditions}

Henceforth we fix a reduction subsystem $\Sigma'$ of $\Sigma^\infty$ such that $`dom\Sigma' = `dom\Sigma^\infty$ as in the preceding section.
As above, $S$ denotes the reduction with respect to $\Sigma'$.
Integrability conditions, investigated in this section, measure the  nonequivalence of various ways of prolongation.

\begin{definition} \rm \label{def:ps}
For every principal derivative $u^k_\mu$ (i.e., $u^k_\mu \in `dom \Sigma^\infty$) we introduce the {\it principal subset} $\mathcal X^k_\mu$ as the set of all monomials $\xi \ne 1$ such that $u^k_{\mu/\xi} \in `dom \Sigma^\infty$. Thus, elements of $\mathcal X^k_\mu$ are principal derivatives strictly lower than $u^k_\mu$ (see Definition~\ref{higher}). 
\end{definition}

\begin{definition} \rm \label{def:IC}
If in $\Sigma$ there is an equation of the form $u^k_\mu = \Phi^k_\mu$ such that the principal subset $\mathcal X^k_\mu$ is nonempty and $\xi \in \mathcal X^k_\mu$, then the condition
$$
\numbered\label{eq:cc1}
\Phi^k_\mu = S D_\xi S u^k_{\mu/\xi}
$$
is called an {\it integrability condition of the first kind\/} at the point $u^k_\mu$.

For every pair $\xi,\eta \in \mathcal X^k_\mu$, the condition
$$
\numbered\label{eq:cc2}
S D_\xi S u^k_{\mu/\xi} = S D_\eta S u^k_{\mu/\eta}
$$
is called an {\it integrability condition of the second kind\/} at the point $u^k_\mu$.
\end{definition}

Let us remind the reader that $\Sigma$ is not necessarily a subset of $\Sigma'$. This explains why integrability conditions of the first kind are needed.

According to Definition~\ref{def:IC}, integrability conditions at the point $u^k_\mu$ are satisfied if any two possible ways of obtaining the value $S u^k_\mu$ lead to one and the same result. 
It is well known that all such integrability conditions follow from a finite subset. 
The main goal of this paper is to reestablish this result in an irredundant way.

For every principal derivative $u^k_\mu$, consider the principal subset $\mathcal X^k_\mu$ ordered by the divisibility relation $\le$.
The following definition is quite standard and reflects properties of the graph of the ordered principal subset.
Let $\approx$ denote the reflexive, symmetric, and transitive closure of the ordering $\le$.
In other words, $p \approx q$ if and only if there exists a finite sequence of monomials $z_r,\dots,z_{2s+1} \in \mathcal X^k_\mu$ such that $p = z_1$, $q = z_{2s+1}$ and $z_{2j-1} \le z_{2j}$ whereas $z_{2j} \ge z_{2j+1}$ for every $j = 1,\dots,s$.
Since $\approx$ is an equivalence relation, we have got a partition $\mathcal X^k_\mu/{\approx}$ of the set $\mathcal X^k_\mu$ into equivalence classes $[x]_\approx$, $x \in \mathcal X^k_\mu$.

\begin{definition} \rm \label{def:conn}
A pair of elements $p,q$ of the principal subset $\mathcal X^k_\mu$ is said to be {\it connected\/} if $p \approx q$. 
Equivalence classes with respect to $\approx$ are called {\it connected components} of $\mathcal X^k_\mu$.
The principal subset $\mathcal X^k_\mu$ is said to be {\it connected}, if it consists of a single connected component, otherwise it is said to be {\it disconnected}. 
\end{definition}

\begin{construction}
\label{suffcc} \rm
For every $u^k_\mu \in `dom \Sigma$ with nonempty principal subset $\mathcal X^k_\mu$ choose one integrability condition of the first kind {\eqref{eq:cc1}}, 
$$
\Phi^k_\mu = S D_\xi S u^k_{\mu/\xi},
$$
where $\xi \in \mathcal X^k_\mu$ is arbitrary.
For every $u^k_\mu \in `dom \Sigma^\infty$ such that the principal subset $\mathcal X^k_\mu$ consists of $s$ connected components $[\xi_1]_\approx, \dots, [\xi_s]_\approx$ with $s \gt 1$, choose arbitrary representatives $\xi_1,\dots,\xi_s$ of these components and consider integrability conditions of the second kind {\eqref{eq:cc2}} in the form of a chain of equations
$$
\numbered\label{eq:cc2*}
S D_{\xi_1} S u^k_{\mu/{\xi_1}} = S D_{\xi_2} S u^k_{\mu/{\xi_2}}
 = \dots = S D_{\xi_s} S u^k_{\mu/{\xi_s}}.
$$
The set of integrability conditions obtained in this way will be called a {\it sufficient set of integrability conditions}.
\end{construction}

Clearly, each $\xi$ of Construction~\ref{suffcc} can be chosen so that $u^k_{\mu/\xi}$ is minimal in $\mathcal X^k_\mu$, hence belongs to $`dom\Sigma$ so that we can replace $S D_\xi S u^k_{\mu/\xi}$ with $S D_\xi \Phi^k_{\mu/\xi}$ to obtain conventional integrability conditions in the sense of the following definition.

\begin{definition} \rm
\label{def:cIC}
Assuming $u^k_{\mu/\xi},u^k_{\mu/\xi_i} \in `dom\Sigma$, integrability conditions of the form
$$
\Phi^k_\mu = S D_\xi \Phi^k_{\mu/\xi}
\quad\text{or}\quad
S D_{\xi_1} \Phi^k_{\mu/\xi_1} = \dots = S D_{\xi_s} \Phi^k_{\mu/\xi_s}
$$
are said to be {\it conventional}.
\end{definition}

\begin{remark}
\label{rem:cIC}
Obviously, every integrability condition can become conventional at the cost of enlarging the system $\Sigma$ by appropriate reduced equations from $\Sigma^\infty$.
\end{remark}

Our immediate goal now is to show that every sufficient set resulting from Construction~\ref{suffcc} implies all the other integrability conditions of Definition~\ref{def:IC}. The following lemma is the key. 
Let $`var F$ denote the finite set of all variables (independent variables and derivatives) a smooth function $F$ depends on.

\begin{lemma}
\label{lemma:cc0}
Let $x$ be an independent variable, $\sigma \in \mathcal X^*$ a monomial, and $F$ a function of independent variables and parametric derivatives.
Let $S D_\tau S D_x p = S D_{x \tau} p$ for every derivative $p \in `var F$ and every monomial $\tau \le \sigma$.
Then $S D_\sigma S D_x F = S D_{x \sigma} F$.
\end{lemma}

\begin{proof}
We have 
$$
D_\sigma (F G) = \sum_{\rho\tau = \sigma} c^{\rho\tau}_\sigma D_\rho F \cdot D_\tau G,
$$
for suitable constants $c^{\rho\tau}_\sigma$.  
Applying $S D_\sigma$ to
$$
D_x F = \sum_{q \in `var F} \frac{\partial F}{\partial q} D_x q, \qquad
S D_x F = \sum_{q \in `var F} \frac{\partial F}{\partial q} S D_x q,
$$
we get
$$
S D_\sigma D_x F 
 = \sum_{q \in `var F} \sum_{\rho\tau = \sigma} c^{\rho\tau}_\sigma S D_\rho (\frac{\partial F}{\partial q}) \cdot S D_{\tau x} q,
$$
whereas
$$
S D_\sigma S D_x F 
 = \sum_{q \in `var F} \sum_{\rho\tau = \sigma} c^{\rho\tau}_\sigma S D_\rho (\frac{\partial F}{\partial q}) \cdot S D_\tau S D_x q.
$$
These two expressions coincide since $S D_\tau S D_x q = S D_{\tau x} q$ holds by assumption for all $q$ and $\tau \le \sigma$.
\end{proof}

\begin{theorem}
\label{th:pass}
Suppose that the reduction subsystem $\Sigma'$ (see Sect.~\ref{sect:rs}) satisfies some sufficient set of integrability conditions as in Construction~\ref{suffcc}. 
Then 
\paritem{(i)}
for all $u^k_\mu \in `dom \Sigma^\infty$ and all $\xi < \mu$ we have
$$
\numbered\label{eq:pass}
S u^k_\mu = S D_\xi S u^k_{\mu/\xi};
$$
\paritem{(ii)}
all integrability conditions in the sense of Definition~\ref{def:IC} hold true
(meaning that $\Sigma$ is passive);
\paritem{(iii)}
for every monomial $\xi$ and every smooth function $f$ on $J^\infty$ we have
$$
S D_\xi f = S D_\xi S f;
$$
\paritem{(iv)}
manifolds $\mathcal E_{\Sigma^\infty}$ and $\mathcal E_{\Sigma'}$ coincide.
\end{theorem}

\begin{proof}
To prove (i) we proceed by induction with respect to $p = u^k_\mu$.
If $u^k_{\mu/\xi}$ exists and is parametric, then {\eqref{eq:pass}} is satisfied trivially since $Su^k_{\mu/\xi} = u^k_{\mu/\xi}$.
It remains to deal with the case when $u^k_{\mu/\xi}$ exists and is principal, i.e., the case of $\xi \in \mathcal X^k_\mu$.
To start with, we consider $p = u^k_\mu$ minimal with respect to the ordering $\prec$. Then \eqref{eq:pass} holds true in a trivial way, since $\mathcal X^k_\mu = \emptyset$ in that case.

To perform the induction step, let us consider an arbitrary derivative $p = u^k_\mu$ assuming validity of {\eqref{eq:pass}} for all $q \prec p$. 
We shall prove 
$$
\numbered\label{equiv}
S D_\sigma S u^k_{\mu/\sigma} = S D_\rho S u^k_{\mu/\rho} 
$$
for all $\sigma,\rho \in \mathcal X^k_\mu$.
We start with the case of $\sigma,\rho$ belonging to one connected component, i.e., $\sigma \approx \rho$. Obviously, this case can be reduced to $\rho \le \sigma$ by the definition of $\approx$. But then it can be further reduced to $\rho \vartriangleleft \sigma$ meaning that $\rho \le \sigma$ and $\sigma/\rho$ is a variable, since $\le$ is the reflexive transitive closure of $\vartriangleleft$ in $\mathcal X^k_\mu$.
Therefore, let $x$ be an independent variable $x$ such that $\rho = x \sigma$.
To prove that $\sigma \equiv \rho$, we establish equalities
$$
S D_\sigma S u^k_{\mu/\sigma} = S D_\sigma S D_x S u^k_{\mu/x\sigma}
 = S D_{x\sigma} S u^k_{\mu/x\sigma}
 = S D_\rho S u^k_{\mu/\rho}.
$$
The first equality follows from $S u^k_{\mu/\sigma} = S D_x S u^k_{\mu/x\sigma}$, which is \eqref{eq:pass} for $\mu/\sigma \prec \mu$ and $\xi = x$, therefore holds by induction assumption. 

To prove the second equality we apply Lemma~\ref{lemma:cc0} to $F = S u^k_{\mu/x\sigma}$.
Let us verify the assumptions. Consider an arbitrary monomial $\tau \le \sigma$ and $q \in `var F$. Then $q \prec u^k_{\mu/x\sigma}$, whence $D_{x\tau} q \prec D_{x\tau} u^k_{\mu/x\sigma} = u^k_{\mu\tau/\sigma} \preceq u^k_{\mu\sigma/\sigma} = u^k_\mu$.
By induction assumption, \eqref{eq:pass} holds with $u^k_\mu$ replaced with $D_{x\tau} q$ and $\xi$ with $x$, meaning that $S D_{x\tau} q = S D_x S D_\tau q$.
Having verified assumptions of Lemma~\ref{lemma:cc0}, we have the second equality.
The third equality is obvious.
Thus, \eqref{equiv} holds for arbitrary $\rho,\sigma$ belonging to one connected component of $\mathcal X^k_\mu$. 

We are left with the case when $\sigma,\rho$ belong to different components.
But if the set $\mathfrak I$ of integrability conditions is sufficient in the sense of Construction~\ref{suffcc}, as it is supposed to be, then $\mathfrak I$ contains an integrability condition $S D_\sigma S u^k_{\mu/\sigma'} = S D_\rho S u^k_{\mu/\rho'}$ with some other $\rho' \approx \rho$ and $\sigma' \approx \sigma$ from the same components, and then \eqref{equiv} holds for $\sigma',\rho'$ by assumption and then for $\sigma,\rho$ by transitivity. 

This means that we have one and the same value $S D_\sigma S u^k_{\mu/\sigma} = S D_\rho S u^k_{\mu/\rho}$ for all $\sigma,\rho \in \mathcal X^k_\mu$. 
To establish \eqref{eq:pass}, it remains to show that this common value is also equal to $S u^k_\mu$. 
If $u^k_\mu \not\in `dom\Sigma$, then $S u^k_\mu = S D_\xi S u^k_{\mu/\xi}$ for some $\xi \in \mathcal X^k_\mu$ by construction of the reduction system $\Sigma'$ (Construction~\ref{1-1sub}). 
If $u^k_\mu \in `dom\Sigma$ and $\mathcal X^k_\mu = \emptyset$, then \eqref{eq:pass} is void.
If $u^k_\mu \in `dom\Sigma$ and $\mathcal X^k_\mu \ne \emptyset$ and $S u^k_\mu = \Phi^k_\mu$, then the sufficient system involves an integrability condition of the first kind $\Phi^k_\mu = S D_\xi S u^k_{\mu/\xi}$ for some $\xi \in \mathcal X^k_\mu$.
Finally, if $u^k_\mu \in `dom\Sigma$ and $\mathcal X^k_\mu \ne \emptyset$ and $S u^k_\mu \ne \Phi^k_\mu$, then \eqref{eq:pass} follows from the same Construction~\ref{1-1sub} again. 
Thus, statement (i) is proved.

Statement~(ii) follows immediately from (i) or~\eqref{equiv}.
Statement~(iii) holds for all functions $f$ if it holds for all derivatives $u^k_\nu$, and then it follows from~(i).
Finally, by~(ii) every equation from system $\Sigma^\infty$ becomes an identity when reduced with respect to $S$. This means that systems $\Sigma^\infty$ and $\Sigma'$ are equivalent (follow one from another). Hence statement~(iv).
\end{proof}

\begin{remark} \rm 
Operators $S D_x S$ on the full jet space $J^\infty$ are $\mathbb R$-linear and satisfy the Leibniz rule, hence they are vector fields.
Moreover, they commute, since $[S D_x S, S D_y S] f
 = S D_x S D_y S f - S D_y S D_x S f 
 = S D_x D_y S f - S D_y D_x S f
 = 0$ by Theorem~\ref{th:pass}(iii). 
Hence the full jet space $J^\infty$ equipped with vector fields $S D_x S$, $x \in \mathcal X$, is a diffiety in the sense of~\cite{B-C-D-K-K-S-T-V-V}. 
Now, by Theorem~\ref{th:pass}(iv) we have 
$C^\infty \mathcal E_{\Sigma^\infty}
 = C^\infty \mathcal E_{\Sigma'}
 \cong C^\infty J^\infty/`Ker S
 \cong S C^\infty J^\infty.
$
Hence $S D_x S$ induce well defined operators on manifold $\mathcal E_{\Sigma^\infty}$ as well, turning it into a diffiety.
\end{remark}

In an attempt to convey the sense of the method we conclude this section with a simple example in dimension two.
For less trivial examples see Section~\ref{sect:ex}.

\begin{example} \rm
\label{ex:1a}
Consider the following system $\Sigma$:
$$
u_{xyyyy} = e, \qquad
u_{xxyyy} = f, \qquad
u_{xxxyy} = g, \qquad 
u_{xxxxy} = h,
$$ 
where $e,f,g,h$ are arbitrary functions of parametric derivatives. Figure~\ref{fig:1a} shows the ordered set $`dom\Sigma^\infty$ of principal derivatives placed within a coordinate system. Symbols $e,f,g,h$ denote the four  generating derivatives $u_{xyyyy},\ u_{xxyyy},\ u_{xxxyy},\ u_{xxxxy} \in `dom\Sigma$, respectively.
The bold dot at $u_{xxxxyyyy}$ denotes a typical principal derivative.
Thick lines show the principal subset $\mathcal X_{xxxxyyyy}$, which is obviously connected. 
Actually, one easily sees that {\it all\/} principal subsets are connected except 
$$
\mathcal X_{xxxxyy} = \{ u_{xxxxy}, u_{xxxyy} \}, \\
\mathcal X_{xxxyyy} = \{ u_{xxxyy}, u_{xxyyy} \}, \\
\mathcal X_{xxyyyy} = \{ u_{xxyyy}, u_{xyyyy} \},
$$ 
which consist of two isolated points each.
Correspondingly, each of the derivatives $u_{xxxxyy}$, $u_{xxxyyy}$, $u_{xxyyyy}$ harbours one nontrivial integrability condition.
They are, respectively,
$$
S D_x e = S D_y f, \quad
S D_x f = S D_y g, \quad\text{and}\quad
S D_x g = S D_y h.
$$

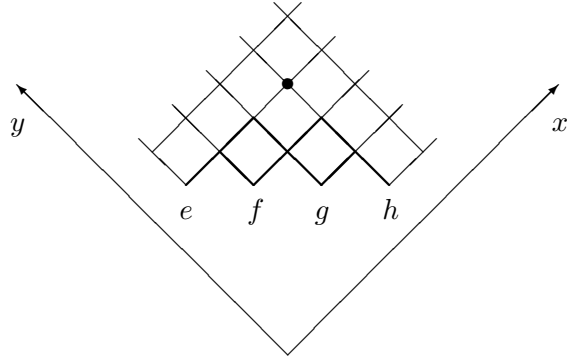
\begin{figure}
\caption{A typical principal subset (Example~\ref{ex:1a})}
\label{fig:1a}
\begin{center}
\unitlength 9mm
\begin{picture}(0,6)(0,0)
\put(0,0){\vector(1,1){4}}
\put(0,0){\vector(-1,1){4}}
\put(-4.1,3.3){\mbox{$y$}}
\put(3.9,3.3){\mbox{$x$}}
\put(-1.5,2.5){\line(-1,1){.7}}
\put(-0.5,2.5){\line(-1,1){1.2}}
\put(0.5,2.5){\line(-1,1){1.7}}
\put(1.5,2.5){\line(-1,1){2.2}}
\put(2,3){\line(-1,1){2.2}}
\put(-1.5,2.5){\line(1,1){2.2}}
\put(-0.5,2.5){\line(1,1){1.7}}
\put(0.5,2.5){\line(1,1){1.2}}
\put(1.5,2.5){\line(1,1){.7}}
\put(-2,3){\line(1,1){2.2}}
\put(-1.6,2){\mbox{$e$}}
\put(-0.6,2){\mbox{$f$}}
\put(0.4,2){\mbox{$g$}}
\put(1.4,2){\mbox{$h$}}
\put(0,4)\picbdot
\thicklines
\put(-1.5,2.5){\line(1,1){1}}
\put(-0.5,2.5){\line(-1,1){.5}}
\put(-0.5,2.5){\line(1,1){1}}
\put(0.5,2.5){\line(-1,1){1}}
\put(0.5,2.5){\line(1,1){.5}}
\put(1.5,2.5){\line(-1,1){1}}
\end{picture}
\end{center}
\end{figure}
\end{example}

\section{Cross-derivatives}
\label{sect:card}

In this section we find an alternative description of the sufficient set suitable for effective implementation.  

Construction~\ref{suffcc} leaves us with the problem to find all principal derivatives with disconnected principal subset. Indeed, a nontrivial integrability condition of the second kind at a point $u^k_\mu$ exists if and only if there are at least two distinct connected components in $\mathcal X^k_\mu$. 
To extend this line of argument further, let us consider different possible descriptions of the quotient sets $\mathcal X^k_\mu/{\approx}$. 

Let $B$ be a subset of $\mathcal X^k_\mu$ such that every element of $\mathcal X^k_\mu$ is connected to an element of $B$ in the sense of Definition~\ref{def:conn}.
Then, obviously, every connected component intersects with $B$. 
In particular, the quotient set $\mathcal X^k_\mu/{\approx}$ is the same as the quotient set $B/{\approx_B}$, where $\approx_B$ is the equivalence relation on $B$ inherited from the relation $\approx$ on $\mathcal X^k_\mu$.
There are two natural choices for $B$, which lead to two alternative descriptions of $\mathcal X^k_\mu/{\approx}$:

\begin{itemize}
\item[(a)] $`min \mathcal X^k_\mu$ = the subset of minimal elements in $\mathcal X^k_\mu$; 
\item[(b)] $`max \mathcal X^k_\mu$ = the subset of maximal elements in $\mathcal X^k_\mu$. 
\end{itemize}
Let $N^{(k)}$ denote the set of all monomials $\mu$ such that $u^k_\mu \in `dom\Sigma$. Assuming $N^{(k)}$ ordered by divisibility, let $M^{(k)} = \min N^{(k)}$ denote the set of all minimal elements in $N^{(k)}$.
Obviously, the subset $`min \mathcal X^k_\mu$ coincides with the intersection $\mathcal X^k_\mu \cap `min N^{(k)} = \mathcal X^k_\mu \cap M^{(k)}$. Define a reflexive and symmetric relation $\uparrow$ on $`min \mathcal X^k_\mu$ by $p \uparrow q$ if $`lcm(p,q) \in \mathcal X^k_\mu$, i.e., if $`lcm(p,q)$ is a proper divisor of $u^k_\mu$. 

Elements of $`max \mathcal X^k_\mu$ are quotients $\mu/x$ with $x \in \mathcal X$ an independent variable such that the derivative $u^k_{\mu/x}$ is principal. To simplify reasoning, we identify $`max\mathcal X^k_\mu$ with a subset of $\mathcal X$. Define a reflexive and symmetric relation $\downarrow$ on $`max \mathcal X^k_\mu \subseteq \mathcal X$ by $x \downarrow y$ if there exists $\sigma \in N^{(k)}$ (equivalently, $\sigma \in M^{(k)}$) and $\sigma \le \mu/x, \mu/y$. The same relation $\downarrow$ can be defined by $x \downarrow y$ if $x = y$ or the derivative $u^k_{\mu/xy}$ is principal as well.

We have the following obvious lemma.

\begin{lemma}
\label{inherited}
The inherited equivalence relation $\approx_{`min \mathcal X^k_\mu}$ coincides with the transitive closure $\uparrow^*$ of\/ $\uparrow$.
The inherited equivalence relation $\approx_{`max \mathcal X^k_\mu}$ coincides with the transitive closure $\downarrow^*$ of\/ $\downarrow$.
\end{lemma}

\begin{corollary} 
\label{col:inherited}
We have bijections
$$
\mathcal X^k_\mu/{\approx} \leftrightarrow `min \mathcal X^k_\mu/{\uparrow^*}
  \leftrightarrow `max \mathcal X^k_\mu/{\downarrow^*}.
$$
\end{corollary}

\begin{proposition} 
\label{prop:lcm}
Let $u^k_\mu$ be a principal derivative such that the principal subset\/ $\mathcal X^k_\mu$ contains nonequivalent elements $\sigma \not\approx \tau$. 
Then $\mu = `lcm(\sigma,\tau)$. Elements $\sigma,\tau$ can be chosen lying in $`min\mathcal X^k_\mu$. 
\end{proposition}

\begin{proof}
Since $\sigma,\tau \in \mathcal X^k_\mu$, we have $`lcm(\sigma,\tau) \le \mu$. If $`lcm(\sigma,\tau) \lt \mu$, then obviously $\sigma \approx \tau$, contradicting the assumptions. The last statement follows from the fact that every connected component intersects with $`min\mathcal X^k_\mu$. 
\end{proof}

Now we introduce cross-derivatives as the ``least common derivatives.''

\begin{definition} \rm
A {\it cross-derivative} is a derivative $u^k_{\LCM(\sigma,\tau)}$, where $u^k_\sigma,u^k_\tau \in `dom\Sigma$ and $\sigma,\tau$ do not divide one another.
\end{definition}

By Proposition~\ref{prop:lcm},  nontrivial integrability conditions of the second kind can be found only at cross derivatives. Hence the well-known result that the number of such integrability conditions is always finite and less or equal to $\frac 12 p(p - 1)$, where $p$ is the number of equations in the system $\Sigma$.

Of course, a cross-derivative gives rise to integrability conditions~\eqref{eq:cc2*} if and only if it satisfies the following nontriviality condition:

\begin{definition} \rm
\label{triv}
A cross-derivative $u^k_\mu$ is said to be {\it trivial\/} if the principal subset $\mathcal X^k_\mu$ is connected. Otherwise it is called {\it nontrivial}.
\end{definition}

\begin{example} \rm
\label{ex:1b}
Generalizing Example~\ref{ex:1a}, consider an arbitrary system of $r$ equations in two dimensions such that $`dom\Sigma$ consists of incomparable derivatives (with respect to $\le$ of definition~\ref{higher}). It is an easy exercise to show that of the $r(r - 1)/2$ cross-derivatives only $r - 1$ are nontrivial.
\end{example}

Let us finish this section with some remarks concerning visualization of the relation $\downarrow$.
The monoid $\mathcal X^*$ can be visualized as the $n$-dimensional grid $\mathbb N^n \subset \mathbb R^n$, where $\mathbb N = \{0,1,2,\dots\}$, via the correspondence $x_1^{r_1} \dots x_n^{r_n} \leftrightarrow (r_1,\dots,r_n)$.
Given a monomial $\mu \in \mathcal X^*$ the {\it cone} generated by $\mu$ is defined to be $C(\mu) = \{ \mu\nu \mid \nu \in \mathcal X^* \}$. A union of cones in $\mathcal X^*$ is called a {\it monomial ideal} (see, e.g.,~\cite{M-S}). 
For each $k$, we have
$$
\{\sigma \in \mathcal X^* \mid u^k_\sigma \in `dom\Sigma^\infty\} = \bigcup_{u^k_\mu \in `dom\Sigma} C(\mu).
$$ 
Hence, to every infinitely prolonged system $\Sigma^\infty$ there corresponds a collection of monomial ideals, one for each $k$, consisting of principal derivatives $u^k_\mu$ with one and the same~$k$.

Monomial ideals are usually visualized by staircase diagrams in $\mathbb R^n$. 
In $\mathbb R^n$, every point $(z_1,\dots,z_n) \in \mathbb N^n$ generates the {\it corner} 
$$
C(z_1,\dots,z_n) = \{ (x_1,\dots,x_n) \in \mathbb R^n \mid \text{ $z_i \le x_i$ for all $i$}\}.
$$
A union of corners, which is an unbounded orthogonal (usually non-convex) polytope with vertices in integer points $\mathbb N^n \subset \mathbb R^n$, is called a {\it staircase diagram}. 
An oriented edge between two integer points $p = (z_1,\dots,z_i,\dots,z_n)$ and $q = (z_1,\dots,\ z_i+1,\ \ldots,z_n)$ is called a {\it direction} from $p$ to $q$. 
A square bounded by four adjacent edges is called a {\it tile}. An $xy$-{\it tile} is a tile parallel to the $xy$-plane. Two of the bounding directions end in a common point, which will be called the {\it vertex} of the tile.
Now, on the staircase diagram $u^k_\mu$ lies in, $`max\mathcal X^k_\mu$ can be seen as the set of all directions that lead to $\mu$. 
Two distinct directions $x,y \in \max\mathcal X$ then satisfy $x \downarrow y$ if and only if the staircase diagram contains the $xy$-tile with the vertex~$\mu$.

\section{The algorithm}
\label{sect:alg}

Before proceeding to more substantial examples, let us finally present the procedure to find a sufficient set of integrability conditions. 
Below the symbol $\#$ denotes the number of elements in a finite set and $`var(\rho) = \{ x \in \mathcal X \mid x \text{ divides } \rho \}$ for $\rho \in \mathcal X^*$ is the set of all variables to occur in a monomial. 
For clarity, we present two separate algorithms, one for integrability conditions of each kind. Both algorithms share partitioning of the system $\Sigma$ into subsystems $\Sigma^{(k)}$ such that $`dom\Sigma^{(k)}$ contains derivatives $u^k_\mu$ of $u^k$. 
Sets $`dom\Sigma^{(k)}$ being denoted by $N^{(k)}$, their minimal elements are then collected in subsets $M^{(k)} \subseteq N^{(k)}$. Of course, the algorithms can share the $k$ loop.

Algorithm~\ref{alg:ic1} to compute integrability conditions of the first kind is very simple. To each non-minimal element $u^k_\sigma \in N^{(k)} \setminus M^{(k)}$ there corresponds exactly one integrability condition of the first kind.

\begin{algorithm}
\caption{Integrability conditions of the first kind}
\label{alg:ic1}
{\it Input\/}: $\Sigma$.
\\
{\it Output\/}: the set $`IC_{\Sigma}^{(1)}$ of integrability conditions of the first kind. 

\begin{algorithmic}[1] 
  \STATE $`IC_\Sigma^{(1)} := \emptyset$ \\
  \FORALL {$k$}
    \STATE $N^{(k)} := \{ \nu \in \mathcal X^* \mid u^k_\nu \in `dom\Sigma \}$ \\
    \STATE $M^{(k)} :=$ the set of minimal elements in $N^{(k)}$
      with respect to $\le$ \\
    \FOR {$\mu \in N^{(k)} \setminus M^{(k)}$} 
      \STATE select arbitrary $\nu \in M^{(k)}$ such that $\nu \lt \mu$ \\
      \STATE adjoin expression $\Phi^k_\mu - D_{\mu/\nu} \Phi^k_\nu$
        to $`IC_\Sigma^{(1)}$ 
    \ENDFOR 
  \ENDFOR
  \RETURN $`IC_\Sigma^{(1)}$
\end{algorithmic}
\end{algorithm}

Before explaining Algorithm~\ref{alg:ic2} to compute integrability conditions of the second kind, let us consider various descriptions offered by Corollary~\ref{col:inherited}. There is no upper bound for the size $\# `min\mathcal X^k_\mu$ since the number of equations in the system $\Sigma$ can be arbitrary, whereas $\#`max\mathcal X^k_\mu$ is bounded by the number of independent variables $\#\mathcal X$. During completion, $\#`min\mathcal X^k_\mu$ grows as new equations are added to the system, while $\#`max\mathcal X^k_\mu$ remains essentially stable.
This is why Algorithm~\ref{alg:ic2} uses the relation $\downarrow$ on $`max\mathcal X^k_\mu$ rather than the relation $\uparrow$ on $`min\mathcal X^k_\mu$. 

\begin{algorithm} 
\caption{Integrability conditions of the second kind}
\label{alg:ic2}
{\it Input\/}: $\Sigma$.
{\it Output\/}: the set $`IC_\Sigma^{(2)}$ of integrability conditions of the second kind 

\begin{algorithmic}[1] \label{alg:ic2}
  \STATE $`IC_\Sigma^{(2)} := \emptyset$ \\
  \FORALL {$k$} 
    \STATE let $N^{(k)},M^{(k)}$ be those of Algorithm~\ref{alg:ic1} \\
    \STATE $L^{(k)} := \{`lcm(\nu_1,\nu_2) \mid
      \{\nu_1,\nu_2\} \subseteq M^{(k)}$ a two-element subset$\}$ \\
    \FOR {$\mu \in L^{(k)}$} 
      \STATE $\mathfrak M = \{ `var(\mu/\sigma) \mid \sigma \in M^{(k)} 
        \text { and } \sigma \lt \mu \}$
      \STATE $\mathfrak N := $ the set of connected components of the hypergraph
         $(\bigcup\mathfrak M, \mathfrak M)$
      \IF {$\#\mathfrak N \gt 1$}
        \STATE
         select arbitrary $\sigma_1,\dots,\sigma_s \in N^{(k)}$ such that
         $`var(\mu/\sigma_i)$ is a subset of the $i$th connected component,
        \STATE adjoin 
         $D_{\mu/\sigma_2} \Phi^k_{\sigma_2} - D_{\mu/\sigma_1} \Phi^k_{\sigma_1},
         \dots$, 
         $D_{\mu/\sigma_s} \Phi^k_{\sigma_s} - D_{\mu/\sigma_1} \Phi^k_{\sigma_1}$
         to $`IC_\Sigma^{(2)}$
      \ENDIF
    \ENDFOR
  \ENDFOR
  \RETURN $`IC_\Sigma^{(2)}$
\end{algorithmic}
\end{algorithm}

Now, Algorithm~\ref{alg:ic2} works as follows. On line 4 cross-derivatives are computed, using only minimal derivatives collected in $M^{(k)}$. This is justified by Proposition~\ref{prop:lcm} as trivial cross-derivatives do not contribute to $`IC_{\Sigma}^{(2)}$.

To explain lines 6 and 7, observe that the reflexive and symmetric relation $\downarrow$ is conveniently represented as the union of relations $\downarrow_\sigma$, where $\sigma$ runs through $M^{(k)}$ and $\downarrow_\sigma$ is defined by $x \downarrow_\sigma y$ if $x = y$ or $\sigma \le \mu/x, \mu/y$. Obviously, each $\downarrow_\sigma$ is an equivalence relation. Moreover, the associated partition of $\mathcal X \subseteq `max\mathcal X^k_\mu$ has only one nontrivial class, apart from one-element sets, and this class can be identified with $`var(\mu/\sigma)$.
Hence, the partition $\mathfrak N$ corresponding to the transitive hull $\downarrow^*$ is the least partition such that every $`var(\mu/\sigma)$ is a subset of some class of $\mathfrak N$. Computing such a least partition amounts to joining all incident subsets. 
Alternatively, $\mathfrak N$ can be described as the set of connected components of the hypergraph $(`max\mathcal X^k_\mu, \mathfrak M)$, where the set of hyperedges is
$\mathfrak M = \{ `var(\mu/\sigma) \mid \sigma$ in $M^{(k)}\}$ and the set of vertices is $`max\mathcal X^k_\mu = \bigcup\mathfrak M$. 
Known algorithms are capable of labeling connected components of a (hyper)graph in expected time linear in the number of vertices, see~\cite{K-Tar}. 
The time complexity of Algorithm~\ref{alg:ic2} is estimated in Remark~\ref{rem:runtime} below.  
An obvious modification is to interlace lines 7 and 8 and break as soon as $\mathfrak N$ has only one connected component. This will further reduce the average running time.

Of course, time spent on redundancy elimination is secondary in comparison with large variance resulting from different completion strategies.
Vital for selecting the proper completion strategy is knowing the available freedom of choice. 
It is, however, clear that the arbitrary selections made on line 6 of Algorithm~\ref{alg:ic1} and lines 10, 11 of Algorithm~\ref{alg:ic2} exhaust the entire freedom of choice of conventional integrability conditions (Definition~\ref{def:cIC}) relative to Remark~\ref{rem:cIC}. 

Let us discuss the whole completion algorithm now. 
If all the integrability conditions of Theorem~\ref{th:pass} are satisfied (i.e., if they reduce to identities as explained in Sect.~\ref{sect:rs}), then the system is passive and no further steps are needed. 
Otherwise let $\bar\Sigma$ denote the extended system obtained by resolving the non-identical integrability conditions with respect to maximal derivatives (under the same ranking) and adjoining them to $\Sigma$.
The system $\bar\Sigma$ is afterwords subject to the same procedure of selecting a reduction subsystem $\bar\Sigma'$ of the infinite prolongation $\bar\Sigma^\infty$, identifying the integrability conditions, etc.
Obviously, we then have
$\mathcal E_{\Sigma'} \subseteq \mathcal E_{\bar\Sigma'}$ and 
$\bar S \circ S = \bar S = S \circ \bar S$.

Of course, the new integrability conditions resulting from extending $\Sigma$ to $\bar\Sigma$ can be of both first and second kind. At the same time yet unresolved integrability conditions of the second kind can trivialize, which is very easy to detect with Algorithm~\ref{alg:ic2}. This explains why proceeding incrementally (resolving one integrability condition at a time) lowers the overall cost of completion. Moreover, the ratio of trivialized and added integrability conditions can be computed quickly and reliably if knowing the new leading derivative alone.  

Of course, being a part of a completion algorithm, the implementation should keep track of the integrability conditions already satisfied in previous steps. But do the once satisfied integrability conditions of $\Sigma$ continue to hold under the new reduction $\bar S$? Recall the inductive proof of~\eqref{eq:pass} (the core statement of Theorem~\ref{th:pass} the others follow from) for the system $\bar\Sigma$. Assume~(\ref{eq:pass}$_\Sigma$), i.e., 
$S u^k_\mu = S D_\xi S u^k_{\mu/\xi}$,
at $u^k_\mu$, while for all $u^l_\nu \prec u^k_\mu$ assume~(\ref{eq:pass}$_{\bar\Sigma}$), i.e., 
$\bar S u^l_\nu
 = \bar S D_\xi \bar S u^l_{\nu/\xi}$.
By (\ref{eq:pass}$_{\bar\Sigma}$) we have $\bar S D_\xi \bar S F = \bar S D_\xi F$ for arbitrary function $F$ of derivatives that precede $u^k_\mu$, in particular, for
$F = S u^k_{\mu/\xi}$.
But then $\bar S D_\xi \bar S u^k_{\mu/\xi}
 = \bar S D_\xi \bar S S u^k_{\mu/\xi}
 = \bar S D_\xi S u^k_{\mu/\xi}
 = \bar S S D_\xi S u^k_{\mu/\xi}
 = \bar S S u^k_\mu
 = \bar S u^k_\mu$.
Thus, we have (\ref{eq:pass}$_{\bar\Sigma}$) at $u^k_\mu$ as well.

An important question is whether the completion algorithm eventually stops. An affirmative answer easily follows from the Dickson lemma, since new integrability conditions can reside only at points $u^k_\mu$ outside the monomial ideals generated by $`dom\Sigma^{(k)}$.

We finish this section with a remark on autoreduction.

\begin{remark} \rm
\label{Buch-1}
Certain grounds exist for maintaining the reduction subsystem non-autoreduced. For example, let the input system $\Sigma$ contain two equations $u_\mu = Fu$, $u_\nu = Gu$, where $F,G$ are linear differential operators with {\it constant\/} coefficients. Let, moreover, $\mu,\nu$ be relatively prime. Then the corresponding integrability condition of the second kind at $u_{\mu\nu}$ is nontrivial, yet automatically satisfied: $D_\nu Fu = FGu = GFu = D_\mu Gu$ on $\mathcal E_\Sigma^\infty$. (In polynomial elimination theory this case is covered by the so-called first Buchberger criterion~\cite{Buch-1}.) 
Now, the crucial identity $FGu = GFu$ (which only appears in expanded form) is much easier to check before applying any reductions for $u_\sigma \in `var FGu = `var GFu$.
To a lesser extent this is so even if $F,G$ have non-constant coefficients etc.  
\end{remark}

\section{Irredundancy}
\label{sect:irr}

In this section we prove that Construction~\ref{suffcc} produces no redundant integrability condition. By a redundant condition one usually means one that can be safely omitted from the checklist, since it is satisfied automatically whenever all the others are. 
To put it more formally, observe that essentials of Construction~\ref{suffcc} depend only on the set $P = `dom\Sigma$ of derivatives $u^k_\mu$ on the left-hand side of the input system~\eqref{eq:Sigma}, while functions $\Phi^k_\xi$ on the right-hand side play the role of parameters. 
Let us therefore consider the whole class $\mathfrak S_P$ of orthonomic systems $\Sigma$ with fixed set $P = `dom\Sigma$, parametrized by arbitrary functions $\Phi^k_\xi$ subject only to requirements of orthonomicity. Obviously, Construction~\ref{suffcc} provides a set of integrability conditions applicable to every member of the class $\mathfrak S_P$.

\begin{definition} \rm
\label{def:redund}
Consider the class $\mathfrak S_P$ of orthonomic systems $\Sigma$ with a fixed set $P = `dom\Sigma$. Let $\mathcal I$ be a set of integrability conditions of $\mathfrak S_P$.
An integrability condition $I \in \mathcal I$ is said to be {\it redundant} if it is satisfied for every choice of right-hand sides $\Phi^k_\xi$ for which all the other integrability conditions $\mathcal I \setminus \{ I \}$ are satisfied. 
The set $\mathcal I$ is said to be irredundant if it contains no redundant integrability condition.  
\end{definition}

A chain~\eqref{eq:cc2*} is to be considered as a sequence of $s - 1$ integrability conditions,
so that each equality sign determines a separate integrability condition.

\begin{remark} \rm
\label{rem:redund}
Definition~\ref{def:redund} implicitly refers to some functional space $\mathcal S$ to choose the right-hand sides $\Phi^k_\xi$ from. Proof of Proposition~\ref{prop:redund} below only requires that $\mathcal S$ contains all polynomials in the independent variables.
\end{remark}

\begin{proposition} 
\label{prop:redund}
The set\/ $\mathcal I$ of integrability conditions resulting from Construction~\ref{suffcc} is irredundant. 
\end{proposition}

\begin{proof} 
To start with, we assume that all integrability conditions from $\mathcal I$ are conventional (see Definition~\ref{def:cIC}). Let $\Phi^k_\mu = S D_{\mu/\xi} S \Phi^k_\xi \in \mathcal I$ be such an integrability condition of the first kind. 
By assigning $\Phi^k_\mu = 1$ and $\Phi^l_\sigma = 0$ for all $u^l_\sigma \in `dom\Sigma \setminus \{ u^k_\mu \}$ we obtain an orthonomic system obviously satisfying all integrability conditions except $\Phi^k_\mu = 1 \ne 0 = S D_{\mu/\xi} S \Phi^k_\xi$. 

Similarly, consider an arbitrary conventional integrability condition of the second kind from $\mathcal I$, say 
$$
\numbered \label{ir_ic}
S D_{\mu/\xi_1} \Phi^k_{\xi_1} = S D_{\mu/\xi_2} \Phi^k_{\xi_2}
 = \dots = S D_{\mu/\xi_s} \Phi^k_{\xi_s}
$$ 
at $u^k_\mu \in `dom\Sigma^\infty$.
Let $[\xi_1], \dots, [\xi_s]$ denote the corresponding equivalence classes in $\min\mathcal X^k_\mu$. 
Let $1 \le r \lt s$ be an arbitrary integer and $I$ denote the $r$th integrability condition in the chain, i.e.,
$S D_{\mu/\xi_r} \Phi^k_{\xi_r} = S D_{\mu/\xi_{r+1}} \Phi^k_{\xi_{r+1}}$.
By Construction~\ref{suffcc}, $\mathcal I$ contains no more than one integrability condition of the first kind of the form $\Phi^k_\mu = S D_{\mu/\sigma} \Phi^k_\sigma$. If such a $\sigma$ exists, let $\Xi$ denote $[\xi_1] \cup \cdots \cup [\xi_r]$ or $[\xi_{r+1}] \cup \cdots \cup [\xi_s]$ whichever contains $\sigma$. If no such $\sigma$ exists, then let $\Xi$ be one (arbitrarily chosen) of these two sets.

For every monomial $\sigma = x_1^{a_1} \dots x_n^{a_n} \in \mathcal X^*$ we introduce the function of independent variables
$$
F_\sigma(x_1,\dots,x_n) = \frac{x_1^{a_1} \dots x_n^{a_n}} {a_1! \dots a_n!},
$$
which obviously satisfies 
$$
\numbered \label{ir_DF}
D_\tau F_\sigma = \begin{cases} 
 F_{\sigma/\tau} & \text{if $\tau \le \sigma$,} \\
 0 & \text{otherwise}.
\end{cases}
$$
Turning back to our proof,
for every $u^l_\sigma \in `dom\Sigma$ we assign $\Phi^l_\sigma$ according to the following simple rule:
$$
\numbered \label{ir_Ph}
\Phi^l_\sigma = \begin{cases} 
 F_{\mu/\sigma} & \text{if $l = k$
   and $\sigma \in \{\mu\} \cup \Xi$,} \\
 0 & \text{otherwise.}
\end{cases}
$$

Consider an arbitrary integrability condition $\Phi^k_\nu = S D_{\nu/\sigma} \Phi^k_\sigma$, $\sigma \lt \nu$, of the first kind from $\mathcal I$.
The only possibility how the left-hand side can be nonzero is when 
$$
\paritem{(A)}
\sigma \in \{\mu\} \cup \Xi, \quad \nu \le \mu,
$$ 
and then it equals $D_{\nu/\sigma} \Phi^k_\sigma = D_{\nu/\sigma} F_{\mu/\sigma} = F_{\mu/\nu}$.
The only possibility how the right-hand side can be nonzero is when 
$$
\paritem{(B)}
\nu \in \{\mu\} \cup \Xi, 
$$
and then it equals the same $F_{\mu/\nu}$. It remains to be checked that conditions (A) and (B) are equivalent.
Before that we observe that the inequality $\sigma \lt \nu$ implies 
$$
\paritem{(C)} \text{if $\sigma,\nu \lt \mu$, then $\sigma,\nu$ both or neither lie in $\Xi$}.
$$ 
Indeed, under these conditions we have $\sigma \approx \nu$ in $\mathcal X^k_\mu$.
 
Let (A) be true. The case $\nu = \mu$ being trivial, consider $\nu \lt \mu$. Then also $\sigma \lt \mu$ and therefore $\sigma \in \Xi$. But then $\nu \in \Xi$ by (C), giving~(B).  
Conversely, let (B) be true.
If $\nu = \mu$, then $\sigma \in \Xi$ since $\Xi$ was chosen that way. Otherwise $\nu \in \Xi$, but then $\sigma \in \Xi$ by (C) again. Therefore, (A) is true. Thus, we have proved the equivalence (A)~$\Leftrightarrow$ (B) and hence validity of all integrability conditions of the first kind.

Now consider an integrability condition of the second kind from $\mathcal I$, at some $u^l_\nu$. In case of $u^l_\nu = u^k_\mu$ the integrability condition is~\eqref{ir_ic}. However, $S D_{\mu/\xi} \Phi^k_{\xi}$ equals $1$ if $\xi \in \Xi$ and $0$ otherwise, hence all equalities~\eqref{ir_ic} hold except $I$.
Thus we are left with the case of an integrability condition $S D_{\nu/\sigma_1} \Phi^k_{\sigma_1} = \cdots = S D_{\nu/\sigma_t} \Phi^k_{\sigma_t}$ from $\mathcal I$, at some $u^l_\nu \ne u^k_\mu$.  
By \eqref{ir_Ph} and \eqref{ir_DF}, the values $S D_{\nu/\sigma_i} \Phi^k_{\sigma_i}$ to be compared at $u^l_\nu$ are all zero except when 
$$
\paritem{(A$_i$)} 
l = k, \quad \sigma_i \lt \nu \le \mu, \quad \sigma_i \in \Xi, 
$$
and then they are 
$D_{\nu/\sigma_i} \Phi^k_{\sigma_i}
 = D_{\nu/\sigma_i} F_{\mu/\sigma_i}
 = F_{\mu/\nu}$
independently of $i$.
Let us show that conditions (A$_i$) are mutually equivalent. However, if one of (A$_i$) holds, then $l = k$ and $\nu \le \mu$, hence $\nu \lt \mu$ (otherwise $u^l_\nu = u^k_\mu$). 
We have $\sigma_j \lt \nu$ for all $j$ by Construction~\ref{suffcc}, hence $\sigma_1 \approx \cdots \approx \sigma_t$ in $\mathcal X^k_\mu$ (although not in $\mathcal X^k_\nu$). Therefore all $\sigma_j$ belong to $\Xi$. Equivalence of conditions (A$_i$) is thereby established.  

Thus we have proved the proposition in case of conventional integrability conditions. 
But since all $\Phi^l_\sigma$ assigned during the proof were functions of independent variables only, we have simply $S D_\nu \Phi^l_\sigma = D_\nu \Phi^l_\sigma$ for the reduction $S$ of any principal derivative. This means that every integrability condition can be identified with a conventional integrability condition. 
Hence the proposition holds for general integrability conditions as well.
\end{proof}

It remains to compare our definition of redundancy with that used by other authors, notably Rust~\cite{Ru,R-R-W}.
Consider the free abelian algebra $\mathcal A^k_\mu$ over the set of abstract generators of the form $D_{\mu/\xi} \Phi^k_\xi$. The total derivatives $D_x$ act upon the generators, hence upon the whole algebra, in a natural way. An integrability condition can be viewed as a difference $D_{\mu/\xi} \Phi^k_{\xi} - D_{\mu/\eta} \Phi^k_{\eta} \in \mathcal A^k_\mu$ of two generators.
Given a finite set $\mathcal I \subset \mathcal A^k_\mu$ of integrability conditions, another integrability condition $I = D_{\mu/\xi} \Phi^k_{\xi} - D_{\mu/\eta} \Phi^k_{\eta}$ is said to be {\it syzygy redundant} if monomials $\mu_i \le \mu$, integrability conditions $I_i = D_{\mu_i/\xi_i} \Phi^k_{\xi_i} - D_{\mu_i/\eta_i} \Phi^k_{\eta_i} \in \mathcal I$, and integers $c_i \in \mathbb Z$ exist such that 
$$
\numbered\label{eq:syz}
I = \sum_i c_i D_{\mu/\mu_i} I_i
 = \sum_i c_i (D_{\mu/\xi_i} \Phi^k_{\xi_i}
   - D_{\mu/\eta_i} \Phi^k_{\eta_i})
$$
holds in $\mathcal A^k_\mu$. 

It is clear that if $I$ is syzygy redundant, then it is also redundant in the sense of Definition~\ref{def:redund} for all choices of $\mathcal S$ (see Remark~\ref{rem:redund}). Hence the sufficient set of integrability conditions resulting from Construction~\ref{suffcc}, proved to be irredundant when $\mathcal S$ contains all polynomials in independent variables, is also syzygy irredundant.

\section{Examples}
\label{sect:ex}

The first two examples compare our algorithms to Algorithm~9 from Wittkopf's dissertation~\cite{Witt}. Wittkopf's algorithm removes syzygy redundancy nearly completely. Experiments with randomly generated monomial ideals showed that Wittkopf's algorithm can miss $r$ redundant integrability conditions in case of ideals with $4r$ generators, but such instances are rather rare.
A surprise was that Wittkopf's algorithm could be substantially slower.

\begin{example} \rm
\label{ex:WK1}
Consider a system $\Sigma$ of the form
$$
u_{xyz} = f_1, \quad u_{xxz} = f_2, \quad u_{yyz} = f_3, \quad u_{xxyy} = f_4. 
$$
We summarize the work of Algorithm~\ref{alg:ic2} in a table: 
$$
\begin{array}{cccccccc}
\hline
\mu &  \multicolumn 4c{`var(\mu/\sigma),\ \sigma =}
 & `max\mathcal X_\mu/{\approx} & `IC^{(2)} \\
            & x y z & x^2 z & y^2 z & x^2 y^2 & \\
\hline
    x^2 y z &    x  &    y  &       &         & \{x\}, \{y\}
                                                &  D_x f_1 = D_y f_2 \\ 
    x y^2 z &    y  &       &   x   &         & \{x\}, \{y\} 
                                                & D_y f_1 = D_x f_3 \\ 
  x^2 y^2 z &  x y  &    y  &   x   &     z   & \{x,y\}, \{z\}
                                                & D_{xy} f_1 = D_z f_4 \\ 
\hline
\end{array}
$$
The first column lists all possible cross-derivatives $\mu$.
Columns 2--5 correspond to the four derivatives $u_\sigma$ from $`dom\Sigma$. These four columns list variables the monomial $\mu/\sigma$ depends on whenever $\sigma$ divides $\mu$, and contain an empty space when $\mu/\sigma$ is not a monomial. 
The sixth column contains the least partition of $`max\mathcal X_\mu$ generated by the sets occurring in columns 2--5 ($`max\mathcal X_\mu$ is the union of these sets). By results of Section~\ref{sect:card} this partition corresponds to the equivalence relation $\approx$ on $`max\mathcal X_\mu$ inherited from~$\mathcal X_\mu$. 
Algorithm~\ref{alg:ic2} also says how to choose the integrability conditions (we omit the reduction symbol). In the first and second row the only possibility is that given in the last column (when $\mu = x^2 y z$ or $\mu = x y^2 z$, each connected component of $\mathcal X_\mu$ contains a single $\sigma$). Contrary to that, one of the connected components of $\mathcal X_{xxyyz}$ contains three monomials $\sigma$, namely $x y z, x^2 z, y^2 z$. Hence, apart from $D_{xy} f_1 = D_z f_4$ shown in the table, there are two other equivalent ways to write the third integrability condition: $D_{yy} f_2 = D_z f_4$ and $D_{xx} f_3 = D_z f_4$.

The example can be visualized (see end of Sect.~\ref{sect:card}) as follows:

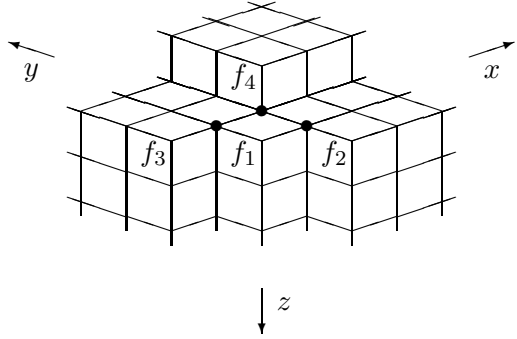
\begin{center}
\begin{picture}(8.6,7.166666666)(-4.3,-4.3)
\put(1.,0){\makebox(0,0){$\bullet$}}
\put(0,.333){\makebox(0,0){$\bullet$}}
\put(-1.,0){\makebox(0,0){$\bullet$}}
\put(-.4,1.1){\makebox(0,0){$f_4$}}
\put(-2.4,-.6){\makebox(0,0){$f_3$}}
\put(1.6,-.6){\makebox(0,0){$f_2$}}
\put(-.4,-.6){\makebox(0,0){$f_1$}}
\put(0,1.33){\line(-3,1){1}}
\put(0,1.33){\line(0,-1){1}}
\put(0,.333){\line(-3,1){1}}
\put(-1.,1.67){\line(-3,1){1.3}}
\put(-1.,1.67){\line(0,-1){1}}
\put(-1.,.667){\line(-3,1){1.3}}
\put(-2.,2.){\line(0,-1){1}}
\put(-2.,-.333){\line(-3,1){1}}
\put(-2.,-.333){\line(0,-1){1}}
\put(-2.,-1.33){\line(-3,1){1}}
\put(-2.,-1.33){\line(0,-1){1.3}}
\put(-2.,-2.33){\line(-3,1){1}}
\put(-3.,0){\line(-3,1){1.3}}
\put(-3.,0){\line(0,-1){1}}
\put(-3.,-1.){\line(-3,1){1.3}}
\put(-3.,-1.){\line(0,-1){1.3}}
\put(-3.,-2.){\line(-3,1){1.3}}
\put(-4.,.333){\line(0,-1){1}}
\put(-4.,-.667){\line(0,-1){1.3}}
\put(2.,-.333){\line(-3,1){1}}
\put(2.,-.333){\line(0,-1){1}}
\put(2.,-1.33){\line(-3,1){1}}
\put(2.,-1.33){\line(0,-1){1.3}}
\put(2.,-2.33){\line(-3,1){1}}
\put(1.,0){\line(-3,1){1}}
\put(1.,0){\line(0,-1){1}}
\put(1.,-1.){\line(0,-1){1.3}}
\put(0,.333){\line(-3,1){1}}
\put(-1.,.667){\line(-3,1){1.3}}
\put(0,-.333){\line(-3,1){1}}
\put(0,-.333){\line(0,-1){1}}
\put(0,-1.33){\line(-3,1){1}}
\put(0,-1.33){\line(0,-1){1.3}}
\put(0,-2.33){\line(-3,1){1}}
\put(-1.,0){\line(-3,1){1}}
\put(-1.,0){\line(0,-1){1}}
\put(-1.,-1.){\line(0,-1){1.3}}
\put(-2.,.333){\line(-3,1){1.3}}
\put(0,1.33){\line(3,1){1}}
\put(0,1.33){\line(0,-1){1}}
\put(0,.333){\line(3,1){1}}
\put(1.,1.67){\line(3,1){1.3}}
\put(1.,1.67){\line(0,-1){1}}
\put(1.,.667){\line(3,1){1.3}}
\put(2.,2.){\line(0,-1){1}}
\put(-2.,-.333){\line(3,1){1}}
\put(-2.,-.333){\line(0,-1){1}}
\put(-2.,-1.33){\line(3,1){1}}
\put(-2.,-1.33){\line(0,-1){1.3}}
\put(-2.,-2.33){\line(3,1){1}}
\put(-1.,0){\line(3,1){1}}
\put(-1.,0){\line(0,-1){1}}
\put(-1.,-1.){\line(0,-1){1.3}}
\put(0,.333){\line(3,1){1}}
\put(1.,.667){\line(3,1){1.3}}
\put(2.,-.333){\line(3,1){1}}
\put(2.,-.333){\line(0,-1){1}}
\put(2.,-1.33){\line(3,1){1}}
\put(2.,-1.33){\line(0,-1){1.3}}
\put(2.,-2.33){\line(3,1){1}}
\put(3.,0){\line(3,1){1.3}}
\put(3.,0){\line(0,-1){1}}
\put(3.,-1.){\line(3,1){1.3}}
\put(3.,-1.){\line(0,-1){1.3}}
\put(3.,-2.){\line(3,1){1.3}}
\put(4.,.333){\line(0,-1){1}}
\put(4.,-.667){\line(0,-1){1.3}}
\put(0,-.333){\line(3,1){1}}
\put(0,-.333){\line(0,-1){1}}
\put(0,-1.33){\line(3,1){1}}
\put(0,-1.33){\line(0,-1){1.3}}
\put(0,-2.33){\line(3,1){1}}
\put(1.,0){\line(3,1){1}}
\put(1.,0){\line(0,-1){1}}
\put(1.,-1.){\line(0,-1){1.3}}
\put(2.,.333){\line(3,1){1.3}}
\put(0,1.33){\line(3,1){1}}
\put(0,1.33){\line(-3,1){1}}
\put(-1.,1.67){\line(3,1){1}}
\put(-1.,1.67){\line(-3,1){1.3}}
\put(-2.,2.){\line(3,1){1}}
\put(1.,1.67){\line(3,1){1.3}}
\put(1.,1.67){\line(-3,1){1}}
\put(0,2.){\line(3,1){1.3}}
\put(0,2.){\line(-3,1){1.3}}
\put(-1.,2.33){\line(3,1){1.3}}
\put(2.,2.){\line(-3,1){1}}
\put(1.,2.33){\line(-3,1){1.3}}
\put(-2.,-.333){\line(3,1){1}}
\put(-2.,-.333){\line(-3,1){1}}
\put(-3.,0){\line(3,1){1}}
\put(-3.,0){\line(-3,1){1.3}}
\put(-4.,.333){\line(3,1){1}}
\put(-1.,0){\line(3,1){1}}
\put(-1.,0){\line(-3,1){1}}
\put(-2.,.333){\line(3,1){1}}
\put(-2.,.333){\line(-3,1){1.3}}
\put(-3.,.667){\line(3,1){1}}
\put(0,.333){\line(3,1){1}}
\put(0,.333){\line(-3,1){1}}
\put(-1.,.667){\line(-3,1){1.3}}
\put(1.,.667){\line(3,1){1.3}}
\put(2.,-.333){\line(3,1){1}}
\put(2.,-.333){\line(-3,1){1}}
\put(1.,0){\line(3,1){1}}
\put(1.,0){\line(-3,1){1}}
\put(0,.333){\line(3,1){1}}
\put(0,.333){\line(-3,1){1}}
\put(-1.,.667){\line(-3,1){1.3}}
\put(3.,0){\line(3,1){1.3}}
\put(3.,0){\line(-3,1){1}}
\put(2.,.333){\line(3,1){1.3}}
\put(2.,.333){\line(-3,1){1}}
\put(1.,.667){\line(3,1){1.3}}
\put(4.,.333){\line(-3,1){1}}
\put(3.,.667){\line(-3,1){1}}
\put(0,-.333){\line(3,1){1}}
\put(0,-.333){\line(-3,1){1}}
\put(-1.,0){\line(3,1){1}}
\put(-1.,0){\line(-3,1){1}}
\put(-2.,.333){\line(3,1){1}}
\put(-2.,.333){\line(-3,1){1.3}}
\put(-3.,.667){\line(3,1){1}}
\put(1.,0){\line(3,1){1}}
\put(1.,0){\line(-3,1){1}}
\put(0,.333){\line(3,1){1}}
\put(0,.333){\line(-3,1){1}}
\put(-1.,.667){\line(-3,1){1.3}}
\put(2.,.333){\line(3,1){1.3}}
\put(2.,.333){\line(-3,1){1}}
\put(1.,.667){\line(3,1){1.3}}
\put(3.,.667){\line(-3,1){1}}
\put(4.6,1.53){\vector(3,1){1}}
\put(5.1,1.20){\makebox(0,0){$x$}}
\put(-4.6,1.53){\vector(-3,1){1}}
\put(-5.1,1.20){\makebox(0,0){$y$}}
\put(0,-3.6){\vector(0,-1){1}}
\put(.5,-3.93){\makebox(0,0){$z$}}
\end{picture}
\end{center}
The reader may wish to locate the tile that induces the equivalence relation $x \approx y$ at $\mu = x^2 y^2 z$.
\end{example}

Example~\ref{ex:WK1} is one of the simplest where Algorithm~9 from Wittkopf's dissertation~\cite{Witt} misses one redundant integrability condition. 
However, Wittkopf's algorithm depends on a choice of what is called compatible ranking~\cite[Def.~10]{Witt} of syzygies, which itself depends on a choice of a ranking for $\Sigma$ (which we fix to be $x \prec y \prec z$) and a permutation of the set $\Sigma$. In Example~\ref{ex:WK1}, Wittkopf's algorithm has a very favorable ratio $\frac{11}{12}$ of correct answers in the set of all $4! = 24$ permutations of $\Sigma$. This ratio can be less favorable in other examples.

\begin{example} \rm
\label{ex:WK2}
Consider a system $\Sigma$ of the form
$$
u_{xxy} = f_1, \quad u_{xxz} = f_2, \quad u_{xyy} = f_3, \\
u_{xzz} = f_4, \quad u_{yyz} = f_5, \quad u_{yzz} = f_6. 
$$
We summarize the work of Algorithm~\ref{alg:ic2} in a table: 
$$
\begin{array}{ccccccccc}
\hline
\mu &  \multicolumn 6c{`var(\mu/\sigma),\ \sigma =}
 & `max\mathcal X_\mu/{\approx} \\
            & x^2 y & x^2 z & x y^2 & y^2 z & x z^2 & y z^2 & `IC^{(2)} \\
\hline
    x^2 y z &   z   &   y   &       &       &       &       & \{y\}, \{z\}
                                                            & D_z f_1 = D_y f_2 \\ 
    x y^2 z &       &       &   z   &   x   &       &       & \{x\}, \{z\}  
                                                            & D_z f_3 = D_x f_4 \\ 
    x y z^2 &       &       &       &       &   y   &   x   & \{x\}, \{y\}  
                                                            & D_y f_5 = D_x f_6 \\ 
    x^2 y^2 &   y   &       &   x   &       &       &       & \{x\}, \{y\}
                                                            & D_y f_1 = D_x f_3 \\ 
    x^2 z^2 &       &   z   &       &       &   x   &       & \{x\}, \{z\}  
                                                            & D_z f_2 = D_x f_5 \\ 
    y^2 z^2 &       &       &       &   z   &       &   y   & \{y\}, \{z\}  
                                                            & D_z f_4 = D_y f_6 \\ 
  x^2 y^2 z &  y z  &   y   &  x z  &   x   &       &       & \{x,y,z\} \\ 
  x^2 y z^2 &   z   &  y z  &       &       &  x y  &   x   & \{x,y,z\} \\ 
  x y^2 z^2 &       &       &   z   &  x z  &   y   &  x y  & \{x,y,z\} \\ 
\hline
\end{array}
$$
For explanation of the table see Example~\ref{ex:WK1}.
The last column shows the unique integrability condition in each of the first six rows and none in the remaining three.
The corresponding diagram is
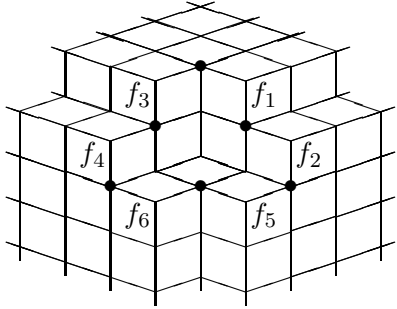
\begin{center}
\begin{picture}(8.6,7.17)(-4.3,-4.3)
\put(0,1.33){\makebox(0,0){$\bullet$}}
\put(1,0){\makebox(0,0){$\bullet$}}
\put(2,-1.33){\makebox(0,0){$\bullet$}}
\put(-1,0){\makebox(0,0){$\bullet$}}
\put(-2,-1.33){\makebox(0,0){$\bullet$}}
\put(0,-1.33){\makebox(0,0){$\bullet$}}
\put(1.4,0.7){\makebox(0,0){$f_1$}}
\put(2.4,-.633){\makebox(0,0){$f_2$}}
\put(-1.4,0.7){\makebox(0,0){$f_3$}}
\put(-2.4,-.633){\makebox(0,0){$f_4$}}
\put(1.4,-1.97){\makebox(0,0){$f_5$}}
\put(-1.4,-1.97){\makebox(0,0){$f_6$}}
\put(1.,1.){\line(-3,1){1}}
\put(1.,1.){\line(0,-1){1}}
\put(1.,0){\line(-3,1){1}}
\put(1.,0){\line(0,-1){1}}
\put(1.,-1.){\line(-3,1){1}}
\put(0,1.33){\line(-3,1){1}}
\put(0,1.33){\line(0,-1){1}}
\put(0,.333){\line(0,-1){1}}
\put(-1.,1.67){\line(-3,1){1.3}}
\put(2.,-.333){\line(-3,1){1}}
\put(2.,-.333){\line(0,-1){1}}
\put(2.,-1.33){\line(-3,1){1}}
\put(2.,-1.33){\line(0,-1){1}}
\put(2.,-2.33){\line(0,-1){1.3}}
\put(1.,0){\line(-3,1){1}}
\put(1.,0){\line(0,-1){1}}
\put(1.,-1.){\line(-3,1){1}}
\put(0,.333){\line(0,-1){1}}
\put(-1.,-1.67){\line(-3,1){1}}
\put(-1.,-1.67){\line(0,-1){1}}
\put(-1.,-2.67){\line(-3,1){1}}
\put(-1.,-2.67){\line(0,-1){1.3}}
\put(-1.,-3.67){\line(-3,1){1}}
\put(-2.,-1.33){\line(-3,1){1}}
\put(-2.,-1.33){\line(0,-1){1}}
\put(-2.,-2.33){\line(-3,1){1}}
\put(-2.,-2.33){\line(0,-1){1.3}}
\put(-2.,-3.33){\line(-3,1){1}}
\put(-3.,-1.){\line(-3,1){1.3}}
\put(-3.,-1.){\line(0,-1){1}}
\put(-3.,-2.){\line(-3,1){1.3}}
\put(-3.,-2.){\line(0,-1){1.3}}
\put(-3.,-3.){\line(-3,1){1.3}}
\put(-4.,-.667){\line(0,-1){1}}
\put(-4.,-1.67){\line(0,-1){1.3}}
\put(1.,-1.67){\line(-3,1){1}}
\put(1.,-1.67){\line(0,-1){1}}
\put(1.,-2.67){\line(-3,1){1}}
\put(1.,-2.67){\line(0,-1){1.3}}
\put(1.,-3.67){\line(-3,1){1}}
\put(0,-1.33){\line(-3,1){1}}
\put(0,-1.33){\line(0,-1){1}}
\put(0,-2.33){\line(0,-1){1.3}}
\put(-2.,-.333){\line(-3,1){1}}
\put(-2.,-.333){\line(0,-1){1}}
\put(-2.,-1.33){\line(-3,1){1}}
\put(-2.,-1.33){\line(0,-1){1}}
\put(-2.,-2.33){\line(-3,1){1}}
\put(-2.,-2.33){\line(0,-1){1.3}}
\put(-2.,-3.33){\line(-3,1){1}}
\put(-3.,0){\line(-3,1){1.3}}
\put(-3.,0){\line(0,-1){1}}
\put(-3.,-1.){\line(-3,1){1.3}}
\put(-3.,-1.){\line(0,-1){1}}
\put(-3.,-2.){\line(-3,1){1.3}}
\put(-3.,-2.){\line(0,-1){1.3}}
\put(-3.,-3.){\line(-3,1){1.3}}
\put(-4.,.333){\line(0,-1){1}}
\put(-4.,-.667){\line(0,-1){1}}
\put(-4.,-1.67){\line(0,-1){1.3}}
\put(-1.,1.){\line(-3,1){1}}
\put(-1.,1.){\line(0,-1){1}}
\put(-1.,0){\line(-3,1){1}}
\put(-1.,0){\line(0,-1){1}}
\put(-2.,1.33){\line(-3,1){1.3}}
\put(-2.,1.33){\line(0,-1){1}}
\put(-2.,.333){\line(-3,1){1.3}}
\put(-3.,1.67){\line(0,-1){1}}
\put(1.,1.){\line(3,1){1}}
\put(1.,1.){\line(0,-1){1}}
\put(1.,0){\line(3,1){1}}
\put(1.,0){\line(0,-1){1}}
\put(2.,1.33){\line(3,1){1.3}}
\put(2.,1.33){\line(0,-1){1}}
\put(2.,.333){\line(3,1){1.3}}
\put(3.,1.67){\line(0,-1){1}}
\put(2.,-.333){\line(3,1){1}}
\put(2.,-.333){\line(0,-1){1}}
\put(2.,-1.33){\line(3,1){1}}
\put(2.,-1.33){\line(0,-1){1}}
\put(2.,-2.33){\line(3,1){1}}
\put(2.,-2.33){\line(0,-1){1.3}}
\put(2.,-3.33){\line(3,1){1}}
\put(3.,0){\line(3,1){1.3}}
\put(3.,0){\line(0,-1){1}}
\put(3.,-1.){\line(3,1){1.3}}
\put(3.,-1.){\line(0,-1){1}}
\put(3.,-2.){\line(3,1){1.3}}
\put(3.,-2.){\line(0,-1){1.3}}
\put(3.,-3.){\line(3,1){1.3}}
\put(4.,.333){\line(0,-1){1}}
\put(4.,-.667){\line(0,-1){1}}
\put(4.,-1.67){\line(0,-1){1.3}}
\put(-1.,-1.67){\line(3,1){1}}
\put(-1.,-1.67){\line(0,-1){1}}
\put(-1.,-2.67){\line(3,1){1}}
\put(-1.,-2.67){\line(0,-1){1.3}}
\put(-1.,-3.67){\line(3,1){1}}
\put(0,-1.33){\line(3,1){1}}
\put(0,-1.33){\line(0,-1){1}}
\put(0,-2.33){\line(0,-1){1.3}}
\put(1.,-1.67){\line(3,1){1}}
\put(1.,-1.67){\line(0,-1){1}}
\put(1.,-2.67){\line(3,1){1}}
\put(1.,-2.67){\line(0,-1){1.3}}
\put(1.,-3.67){\line(3,1){1}}
\put(2.,-1.33){\line(3,1){1}}
\put(2.,-1.33){\line(0,-1){1}}
\put(2.,-2.33){\line(3,1){1}}
\put(2.,-2.33){\line(0,-1){1.3}}
\put(2.,-3.33){\line(3,1){1}}
\put(3.,-1.){\line(3,1){1.3}}
\put(3.,-1.){\line(0,-1){1}}
\put(3.,-2.){\line(3,1){1.3}}
\put(3.,-2.){\line(0,-1){1.3}}
\put(3.,-3.){\line(3,1){1.3}}
\put(4.,-.667){\line(0,-1){1}}
\put(4.,-1.67){\line(0,-1){1.3}}
\put(-2.,-.333){\line(3,1){1}}
\put(-2.,-.333){\line(0,-1){1}}
\put(-2.,-1.33){\line(3,1){1}}
\put(-2.,-1.33){\line(0,-1){1}}
\put(-2.,-2.33){\line(0,-1){1.3}}
\put(-1.,0){\line(3,1){1}}
\put(-1.,0){\line(0,-1){1}}
\put(-1.,-1.){\line(3,1){1}}
\put(0,.333){\line(0,-1){1}}
\put(-1.,1.){\line(3,1){1}}
\put(-1.,1.){\line(0,-1){1}}
\put(-1.,0){\line(3,1){1}}
\put(-1.,0){\line(0,-1){1}}
\put(-1.,-1.){\line(3,1){1}}
\put(0,1.33){\line(3,1){1}}
\put(0,1.33){\line(0,-1){1}}
\put(0,.333){\line(0,-1){1}}
\put(1.,1.67){\line(3,1){1.3}}
\put(1.,1.){\line(3,1){1}}
\put(1.,1.){\line(-3,1){1}}
\put(0,1.33){\line(3,1){1}}
\put(0,1.33){\line(-3,1){1}}
\put(-1.,1.67){\line(3,1){1}}
\put(-1.,1.67){\line(-3,1){1.3}}
\put(-2.,2.){\line(3,1){1}}
\put(2.,1.33){\line(3,1){1.3}}
\put(2.,1.33){\line(-3,1){1}}
\put(1.,1.67){\line(3,1){1.3}}
\put(1.,1.67){\line(-3,1){1}}
\put(0,2.){\line(3,1){1.3}}
\put(0,2.){\line(-3,1){1.3}}
\put(-1.,2.33){\line(3,1){1.3}}
\put(3.,1.67){\line(-3,1){1}}
\put(2.,2.){\line(-3,1){1}}
\put(1.,2.33){\line(-3,1){1.3}}
\put(2.,-.333){\line(3,1){1}}
\put(2.,-.333){\line(-3,1){1}}
\put(1.,0){\line(3,1){1}}
\put(1.,0){\line(-3,1){1}}
\put(3.,0){\line(3,1){1.3}}
\put(3.,0){\line(-3,1){1}}
\put(2.,.333){\line(3,1){1.3}}
\put(4.,.333){\line(-3,1){1}}
\put(-1.,-1.67){\line(3,1){1}}
\put(-1.,-1.67){\line(-3,1){1}}
\put(-2.,-1.33){\line(3,1){1}}
\put(-2.,-1.33){\line(-3,1){1}}
\put(-3.,-1.){\line(-3,1){1.3}}
\put(0,-1.33){\line(3,1){1}}
\put(0,-1.33){\line(-3,1){1}}
\put(-1.,-1.){\line(3,1){1}}
\put(1.,-1.){\line(-3,1){1}}
\put(1.,-1.67){\line(3,1){1}}
\put(1.,-1.67){\line(-3,1){1}}
\put(0,-1.33){\line(3,1){1}}
\put(0,-1.33){\line(-3,1){1}}
\put(-1.,-1.){\line(3,1){1}}
\put(2.,-1.33){\line(3,1){1}}
\put(2.,-1.33){\line(-3,1){1}}
\put(1.,-1.){\line(-3,1){1}}
\put(3.,-1.){\line(3,1){1.3}}
\put(-2.,-.333){\line(3,1){1}}
\put(-2.,-.333){\line(-3,1){1}}
\put(-3.,0){\line(3,1){1}}
\put(-3.,0){\line(-3,1){1.3}}
\put(-4.,.333){\line(3,1){1}}
\put(-1.,0){\line(3,1){1}}
\put(-1.,0){\line(-3,1){1}}
\put(-2.,.333){\line(-3,1){1.3}}
\put(-1.,1.){\line(3,1){1}}
\put(-1.,1.){\line(-3,1){1}}
\put(-2.,1.33){\line(3,1){1}}
\put(-2.,1.33){\line(-3,1){1.3}}
\put(-3.,1.67){\line(3,1){1}}
\put(0,1.33){\line(3,1){1}}
\put(0,1.33){\line(-3,1){1}}
\put(-1.,1.67){\line(3,1){1}}
\put(-1.,1.67){\line(-3,1){1.3}}
\put(-2.,2.){\line(3,1){1}}
\put(1.,1.67){\line(3,1){1.3}}
\put(1.,1.67){\line(-3,1){1}}
\put(0,2.){\line(3,1){1.3}}
\put(0,2.){\line(-3,1){1.3}}
\put(-1.,2.33){\line(3,1){1.3}}
\put(2.,2.){\line(-3,1){1}}
\put(1.,2.33){\line(-3,1){1.3}}
\end{picture} 
\end{center}

Wittkopf's Algorithm~9 gives an incorrect number of integrability conditions (seven) in $540$ of the full number $6! = 720$ of permutations of $\Sigma$. Thus, the ratio of correct answers is only $1/4$ now. 
\end{example}

\begin{remark} \rm
\label{rem:runtime}
In randomly generated examples, Wittkopf's Algorithm~9 ran substantially longer than ours on the same data.
Both algorithms take advantage of the partitioning $\Sigma = \bigcup\Sigma^{(k)}$. Let us therefore attempt comparison in case of one dependent variable (so that there is no $k$ loop).

The outer loop of Wittkopf's algorithm runs over the syzygy system $\mathcal S$, which has $O(r^2)$ elements, where $r = \#\Sigma$. At each run, subset $\mathcal S' \subseteq \mathcal S$ of already executed (accepted or rejected) syzygies is incremented. Processing elements $s' \in \mathcal S'$ in Step 3.1 costs $\#\mathcal S'$ time units. Processing pairs of elements $s',s'' \in \mathcal S'$ in Step 3.2 costs at least $\#\mathcal S'$ time units again ($s'$ and $s''$ are not independent). This suggests running time at least $O(r^4)$.

In our Algorithm~\ref{alg:ic2}, $N^{(k)}$ as well as $M^{(k)}$ have $O(r)$ elements. The main loop $5$--$12$ runs over $L^{(k)}$, which has $O(r^2)$ elements. 
At each run, building $\mathfrak M$ on line 6 requires time proportional to $\#M^{(k)}$. Obtaining connected components of the hypergraph $\mathfrak M$ on line 7 requires time proportional to $\#M^{(k)} + \#\mathcal X$, where typically $\#M^{(k)} \ge \#\mathcal X$. This suggests $O(r^3)$ running time.
\end{remark}

\begin{example} 
\label{ex:Cab26}
\rm
Since the ring $\mathbb R[\partial/\partial x_1,\dots,\partial/\partial x_n]$ of linear differential operators with real coefficients is isomorphic to the polynomial ring $\mathbb R[x_1,\dots,x_n]$, PDE algorithms can be applied to polynomial ideals as well, in which case they compute Gr\"obner bases.

Caboara et al.~\cite{C-K-R} used the syzygy approach to minimize the number of $S$-polynomials when computing a Gr\"obner basis of a polynomial ideal. 
When their Example~26 is translated into the PDE language, the following Janet monomials correspond to maximal derivatives:
$$
\sigma_1 =       x_2^2 x_3^6 x_4   x_5^6, \qquad
\sigma_3 = x_1^8 x_2^2 x_3^6, \\
\sigma_2 = x_1^8 x_2         x_4   x_5^4, \qquad
\sigma_4 = x_1^8       x_3^6   x_5^4.
$$
As summarized in the following table, Algorithm~\ref{alg:ic2} 
reveals rather immediately that one of the four existing cross-derivatives is trivial:
$$
\begin{array}{cccccc}
\hline
\mu &  \multicolumn 4c{\text{subsets }C_\mu(\sigma),\ \sigma =}
 & \mathcal X_\mu/{\approx} \\
      & \sigma_1 & \sigma_2 & \sigma_3 & \sigma_4 \\
\hline
  x_1^8 x_2^2 x_3^6 x_4 x_5^6 & x_1 & x_2 x_3 x_5 & x_4 x_5 & x_2 x_4 x_5 &
    \{x_1\}, \{x_2,x_3,x_4,x_5\} \\ 
  x_1^8 x_2^2 x_3^6 x_4 x_5^4 &   & x_2 x_3 & x_4 x_5 & x_2 x_4 &
    \{x_2,x_3,x_4,x_5\} \\ 
  x_1^8 x_2 x_3^6 x_4 x_5^4 &   & x_3 &   &  x_2 x_4 & \{x_2,x_4\}, \{x_3\} \\ 
  x_1^8 x_2 x_3^6 x_5^4 &   &   & x_5 & x_2 & \{x_2\}, \{x_5\} \\ 
\hline
\end{array}
$$
Thus, we arrive at exactly three nontrivial integrability conditions in full accordance with the result of~\cite[Ex. 26]{C-K-R}.
\end{example}

Finally, we give an example where integrability conditions ordered by divisibility form a chain.

\begin{example} \rm
\label{ex:chain}
Let $n \gt 2$ be arbitrary. Consider the following $n$ Janet monomials in $n$ variables:
$$
x_2 x_3 x_4 \dots x_n, \\
x_1^2 x_3 x_4 \dots x_n, \\
x_1^2 x_2^2 x_4 \dots x_n, \\
\dots, \\
x_1^2 x_2^2 x_3^2 \dots x_{n - 1}^2.
$$
The cross-derivatives 
$$
x_1^2 x_2 x_3 x_4 \dots x_n, \\
x_1^2 x_2^2 x_3 x_4 \dots x_n, \\
x_1^2 x_2^2 x_3^2 x_4 \dots x_n, \\
\dots, \\
x_1^2 x_2^2 x_3^2 \dots x_{n - 1}^2 x_n,
$$
are all nontrivial and form a chain of length $n - 1$.
\end{example}

\section{The poset of nontrivial cross-derivatives}

In this section we explore simplest relations between divisibility properties of cross-derivatives and triviality of the corresponding integrability conditions of the second kind. 
Results concerning different variables $u^k$ being totally independent, it suffices to consider the case of a single variable $u$ and omit the upper index~$k$.

Given an orthonomic system $\Sigma$, let $P_\Sigma$ denote the poset of nontrivial cross-derivatives in the sense of Definition~\ref{triv} under ordering by divisibility.
The following proposition shows that minimal elements of $P_\Sigma$ coincide with minimal cross derivatives of $\Sigma$.

\begin{proposition}
Let $\mu$ be a minimal cross-derivative. Then $\mu$ is nontrivial, meaning that $\mu \in P_\Sigma$. 
\end{proposition}

\begin{proof}
We remove nonminimal elements of $`dom\Sigma$ first to ensure that $`dom\Sigma$ is an antichain without affecting the minimal cross-derivatives.
Let $\mu  = `lcm(\alpha,\beta)$ with $\alpha \ne \beta$ in $`dom\Sigma$. Supposing that $\mu$ is trivial, we arrive at a contradiction. By triviality of $\mu$ we have $\alpha \approx \beta$ in $\mathcal X_\mu$. Hence there exist $\zeta_1,\dots,\zeta_{2s+1} \in \mathcal X_\mu$ such that 
$\alpha \le \zeta_1$, $\zeta_1 \ge \zeta_2$, $\zeta_2 \le \zeta_3,\dots,\zeta_{2s-1} \ge \zeta_{2s}$, $\zeta_{2s} \le \zeta_{2s+1}$, $\zeta_{2s+1} \ge \beta$. Consequently there exist $\sigma_1,\dots,\sigma_{s} \in `dom\Sigma$ such that 
$\zeta_{2i-1},\zeta_{2i},\zeta_{2i+1} \in C(\sigma_i)$. 
Then $`lcm(\alpha,\sigma_1) \le \zeta_1 \lt \mu$, which contradicts minimality of $\mu$ unless $\alpha,\sigma_1$ are comparable. But $`dom\Sigma$ is an antichain, hence $\alpha = \sigma_1$.
By repeating the same argument we get $\alpha = \sigma_1 = \sigma_2 = \dots = \sigma_s = \beta$, contradicting the assumptions on $\alpha,\beta$.
\end{proof}

Hence minimal elements of $P_\Sigma$ coincide with the minimal integrability conditions in Reid's sense~\cite{Re}. It follows that nonminimal and nontrivial integrability conditions in our sense can be interpreted as a solution to Reid's problem of finding an irredundant set of supplementary conditions.

Turning back to the poset~$P_\Sigma$, we show below that it can be an antichain (the generic case; all points are minimal) as well as a chain of length less than the number $n$ of independent variables (a single minimal point).

\begin{lemma}
\label{poset}
Let $\nu \lt \mu$ be two nontrivial cross-derivatives.
Then $\mathcal X_\nu \subseteq \mathcal X_\mu$ lies entirely within an equivalence class of\/ $\approx$ at $u_\mu$. 
\end{lemma}

\begin{proof}
Obviously, $\mathcal X_\nu \subseteq \mathcal X_\mu$.
Let $\xi,\eta \in \mathcal X_\nu$ be arbitrary. 
Since $\xi,\nu,\mu \in \mathcal X_\mu$ and $\xi \lt \nu$, $\eta \lt \nu$, we have $\xi \approx \eta$ at~$u_\mu$ and the lemma follows.
\end{proof}

It easily follows that $P_\Sigma$ does not contain a chain of length $n$, where $n$ is the number of independent variables.  
Yet Example~\ref{ex:chain} shows that $P_\Sigma$ can be a single chain of length $n - 1$.

In case of general position, $P_\Sigma$ is an antichain and all nontrivial cross-derivatives are minimal:

\begin{proposition}
If the generating points of\/ $`dom\Sigma \subseteq \mathbb N^n$ have no coordinate in common, then $P_\Sigma$ is an antichain. 
\end{proposition}

\begin{proof}
Let $\alpha = x_1^{a_1} \dots x_n^{a_n}$, $\beta = x_1^{b_1} \dots x_n^{b_n}$ be such that $a_i \ne b_i$ for all $i = 1,\dots,n$.
Let $\mu = `lcm(\alpha,\beta)$ be nontrivial. However, 
$`var(\mu/\alpha)$ and $`var(\mu/\beta)$ are two disjoint sets whose union is the whole $\mathcal X$, hence they constitute the partition $\mathcal X/{\approx}$.
By Lemma~\ref{poset}, there is no room for a nontrivial cross-derivative above $\mu$.
\end{proof}

Interesting combinatorial questions about~$P_\Sigma$ had to be left aside.
Knowing the number of elements of $P_\Sigma$ can be useful, the more so since there seem to be no similar results under the syzygy approach.
For $\# P_\Sigma$ we have the following exact upper bounds in low dimensions: $\#\Sigma - 1$ when $n = 2$ (obvious) and $3\,\#\Sigma - 6$ when $n = 3$ (the same as the number of edges in a planar graph).
However, computer experiments show that ``average'' numbers are substantially lower. Combinatorial aspects of~$P_\Sigma$ are also a subject of the recent work~\cite{Kap}.

\section*{Acknowledgements}

The author acknowledges the support from GA\v{C}R under grant 201/04/0538 and from the M\v{S}MT under project MSM4781305904.
Special thanks are due, in particular, to V.P. Gerdt for encouragement and enlightening discussions and to anonymous referees of the previous version of this paper.
The author also owes very much to the Special Semester on Gr\"obner Bases supported by RICAM (the Radon Institute for Computational and Applied Mathematics, Austrian Academy of Science, Linz) and organized by RICAM and RISC (Research Institute for Symbolic Computation, Johannes Kepler University, Linz, Austria) under the scientific direction of Professor Bruno Buchberger.


\begin{thebibliography}{99}

\small
\bibitem{A-H}
J. Apel and R. Hemmecke, 
Detecting unnecessary reductions in an involutive basis computation, 
{\it J. Symbolic Comput.} {\bf 40} (2005) 1131--1149


\bibitem{B-C-D-K-K-S-T-V-V}
A.V. Bocharov, V.N. Chetverikov, S.V. Duzhin, N.G. Khor'kova, I.S. Krasil'shchik, A.V. Samokhin, Yu.N. Torkhov, A.M. Verbovetsky and A.M. Vinogradov, 
{\it Symmetries and Conservation Laws for Differential Equations of Mathematical Physics}, Translations of Mathematical Monographs 182 (Amer. Math. Society, Providence, RI, 1999).
 
\bibitem{Bou}
F. Boulier, An optimization of Seidenberg's elimination algorithm in differential algebra, {\it Math. Comput. Simulation} {\bf 42} (1996) 439--448

\bibitem{Buch}
B. Buchberger, Ein Algorithmus zum Auffinden der Basiselemente des Restklassenringes nach einem nulldimensionalen Polynomideal, Ph.D. Thesis, Univ- Innsbruck, 1965.

\bibitem{Buch-1}
B. Buchberger, A criterion for detecting unnecessary reductions in the construction of Gr\"obner bases, in: E.W. Ng, editor, {\it Proc. EUROSAM '79,} Lecture Notes in Computer Science 72 (Springer, Berlin, 1979) 3--21.

\bibitem{C-K-R}
M.~Caboara, M.~Kreuzer and L.~Robbiano,
Efficiently computing minimal sets of critical pairs,
{\it J. Symb. Comp.} {\bf 38} (2004) 1169--1190.


\bibitem{Kap}
S.~Kappes,
Orthogonal surfaces. A combinatorial approach, 
PhD. Thesis, Tech. Univ. Berlin, 2007.


\bibitem{G-M}
R.~Gebauer and M.~M\"oller,
On an instalation of Buchberger's algorithm,
{\it J. Symb. Comp.} {\bf 6} (1988) 275--286.

\bibitem{Ger}
V.P.~Gerdt,
Gr\"obner bases and involutive methods for algebraic and differential equations,
{\it Math. Comput. Modelling} {\bf 25} (1997) 75--90.

\bibitem{G-B1}
V.P.~Gerdt and Yu.A.~Blinkov,
Involutive bases of polynomial ideals,
{\it Math. Comput. Simul.} {\bf 45} (1998) 519--541.





\bibitem{He} W. Hereman, 
Review of symbolic software for Lie symmetry analysis, 
{\it Math. Comput. Modelling} {\bf 25} (1997) 115--132.


\bibitem{Hu} E.~Hubert,
Notes on triangular sets and triangulation-decomposition algorithms II: Differential systems, in: F. Winkler and U. Langer (Eds.), {\it Symbolic and Numerical Scientific Computation}, Proc. Conf. Hagenberg, Austria, 2001, Lecture Notes in Computer Science 2630 (Springer, Berlin, 2003) 40--87.


\bibitem{CRC}
N.H. Ibragimov, ed., {\it CRC Handbook of Lie Group Analysis of Differential Equations}, Vol. III. New Trends in Theoretical Developments and Computational Methods (CRC Press, Boca Raton, 1996).

\bibitem{Jan} M. Janet, 
{\it Le\c{c}ons sur les Syst\`emes d'\`Equations aux Deriv\'ees Partielles} 
(Gauthier-Villars, Paris, 1929). 


\bibitem{Kr-LyI}
B. Kruglikov and V. Lychagin, 
Mayer brackets and solvability of PDEs. I 
{\it Differential Geom. Appl.} {\bf 17} (2002) 251--272. 

\bibitem{Kr-LyII}
B. Kruglikov and V. Lychagin, 
Mayer brackets and solvability of PDEs. II 
{\it Trans. Amer. Math. Soc.} {\bf 358} (2006) 1077--1103.


\bibitem{Kuch}
W. K\"uchlin,
A confluence criterion based on the generalised Newman lemma, in: B.F. Caviness, ed., EUROCAL '85, Proc. Conf. Linz, April 1--3, 1985, Vol. 2, {\it Lecture Notes in Comp. Sci.} {\bf 204} (Springer, Berlin, 1985) 390--399.



\bibitem{GIFT} 
M. Marvan, 
Sufficient set of integrability conditions of an orthonomic system: Extended abstract, in: J. Calmet, W.M. Seiler and R.W. Tucker (eds.), 
{\it Global Integrability of Field Theories}, Proc. GIFT 2006, Cockcroft Inst., Daresbury, November 1--3, 2006 (Universit\"atsverlag, Karlsruhe, 2006), 243--247; 
 {\tt http://www.uvka.de/univerlag/volltexte/2006/164/pdf/Tagungsband\_GIFT.pdf}

\bibitem{M-S}
E. Miller and B. Sturmfels, {\it Combinatorial Commutative Algebra},
Springer Graduate Texts in Math 227 (Springer, New York, 2004).

\bibitem{Pom}
J.F. Pommaret, {\it Systems of Partial Differential Equations and Lie Pseudogroups}, (Gordon and Breach, New York et al., 1978).

\bibitem{Re} G.J. Reid, 
Algorithms for reducing a system of PDEs to standard form, determining the dimension of its solution space and calculating its Taylor series solution,
{\it Eur. J. Appl. Math.} {\bf 2} (1991) 293--318. 

\bibitem{R-L-W}
G.J. Reid, P. Lin and A.D. Wittkopf, 
Differential elimination-completion algorithms for DAE and PDAE, {\it Stud. Appl. Math.} {\bf 106} (2001) 1--45. 

\bibitem{R-W-B}
G.J. Reid, A.D. Wittkopf and A. Boulton, 
Reduction of systems of nonlinear partial differential equations to simplified involutive forms, {\it Eur. J. Appl. Math.} {\bf 7} (1996) 604--635. 

\bibitem{Riq} C. Riquier, 
{\it Les Syst\`emes d'\`Equations aux Deriv\'ees Partielles}, Gauthier-Villars, Paris, 1910. 

\bibitem{Ru}
C.J.~Rust,
Rankings of derivatives for elimination algorithms, and formal solvability of analytic PDE,
Ph.D. Thesis, University of Chicago, 1998.

\bibitem{Ru-R}
C.J.~Rust and G.J.~Reid,
Rankings of partial derivatives, 
in: {\it Proc. ISSAC 1997}
(ACM Press, New York, 1997) 9--16.

\bibitem{R-R-W}
C.J.~Rust, G.J.~Reid and A.D.~Wittkopf,
Existence and uniqueness theorems for formal power series solutions of analytic differential systems, 
in: {\it Proc. ISSAC 1999}
(ACM Press, New York, 1999) 105--112.

\bibitem{Sau}
D.J. Saunders, {\it The Geometry of Jet Bundles}, London Math. Soc. Lect. Notes Series 142 (Cambridge Univ. Press, Cambridge, 1989).

\bibitem{K-Tar}
R.M. Karp and R.E. Tarjan, 
Linear expected-time algorithms for connectivity problems. 
{\it J. Algorithms} {\bf 1} (1980) 374--393.

\bibitem{Th} 
J.M. Thomas, 
Riquier's existence theorems, 
{\it Ann. Math.} {\bf 30} (1928--1929) 285--310. 



\bibitem{Tr}
A. Tresse,
Sur les invariants diff\'erentiels des groupes de transformations,
{\it Acta Math.} {\bf 18} (1894) 1--88. 

\bibitem{V-K-K-M}
A.M. Verbovetsky, I.S. Krasil'shchik, P. Kersten and M. Marvan,
{\it Homological Methods in Geometry of Partial Differential Equations}
(MCCME, Moscow, under preparation) (in Russian).

\bibitem{Win}
F. Winkler, Reducing the complexity of the Knuth--Bendix completion algorithm: a ``unification'' of different approaches, in: B.F. Caviness, ed., EUROCAL '85, Proc. Conf. Linz, April 1--3, 1985, Vol. 2, {\it Lecture Notes in Comp. Sci.} {\bf 204} (Springer, Berlin, 1985) 378--389.

\bibitem{W-Bu}
F. Winkler and B. Buchberger, 
A criterion for eliminating unnecessary reductions in the Knuth--Bendix algorithm, in: J. Demetrovics, G. Katona, A. Salomaa, eds., {\it Algebra, Combinatorics and Logic in Computer Science, Vol. II}, Colloquia Mathematica Societatis J. Bolyai 42 (J. Bolyai Math. Soc. and North-Holland, 1986) 849--869.

\bibitem{W-B-L-R}
F. Winkler, B. Buchberger, F. Lichtenberger and H. Rolletschek,
Algorithm 628: an algorithm for constructing canonical bases of polynomial ideals, {\it ACM Trans. Math. Software} {\bf 11} (1) 66-78.
 
\bibitem{Witt}
A.D. Wittkopf, Algorithms and implementations for differential elimination, PhD. Thesis, Simon Fraser Univ., 2004. 
\end{thebibliography}
\end{document}